\documentclass[]{aa}
\usepackage{graphicx}
\usepackage{txfonts}
\usepackage[colorlinks]{hyperref}
\hypersetup{colorlinks=true,linkcolor=blue,citecolor=blue,filecolor=blue,urlcolor=blue,}
\usepackage{amsmath}
\usepackage{algorithm}
\usepackage{amsfonts}
\usepackage{multirow}
\usepackage{mathrsfs}
\usepackage{comment}
\usepackage{bm}
\usepackage{rotating}
\usepackage{color}
\usepackage{graphicx}
\usepackage{subfigure}
\usepackage{savesym}
\usepackage[flushleft]{threeparttable}
\savesymbol{tablenum}
\restoresymbol{SIX}{tablenum}

\usepackage{lineno}

\newcommand{\cha}{{\it Chandra}\,}

\begin{document}

   \title{The properties of X-ray selected AGN in protoclusters pinpointed by Enormous Lya Nebulae}

    \author{A. Traina \inst{1}, F. Vito \inst{1}, F. Arrigoni-Battaia \inst{2}, C.-C. Chen \inst{3}, C. Vignali \inst{4}, X. Prochaska \inst{5}, S. Cantalupo \inst{6}, A. Pensabene \inst{6}, P. Tozzi \inst{7}, A. Travascio \inst{6}, R. Gilli \inst{1}, M. N. Isla Llave \inst{1,4}, S. Marchesi \inst{4,8,1}, G. Mazzolari \inst{9,1}}
    
   \institute{Istituto Nazionale di Astrofisica (INAF) - Osservatorio di Astrofisica e Scienza dello Spazio (OAS), via Gobetti 101, I-40129 Bologna, Italy
        \and 
            Max-Planck-Institut für Astrophysik, Karl-Schwarzschild-Str. 1, D-85748 Garching, Germany   
        \and
            Academia Sinica Institute of Astronomy and Astrophysics (ASIAA), 11F of Astronomy-Mathematics Building, AS/NTU, No. 1, Section 4, 12 Roosevelt Road, Taipei 106319, Taiwan            
        \and
            Dipartimento di Fisica e Astronomia (DIFA), Universit\`a di Bologna, via Gobetti 93/2, I-40129 Bologna, Italy
        \and 
            Department of Astronomy and Astrophysics, UCO/Lick Observatory, University of California, Santa Cruz, 1156 High Street, Santa Cruz, CA 95064, USA
        \and
            Dipartimento di Fisica “G. Occhialini”, Università degli Studi di Milano-Bicocca, Piazza della Scienza 3, 20126 Milano, Italy            
        \and
            INAF – Osservatorio Astrofisco di Arcetri, Largo E. Fermi 5, 50127 Firenze, Italy            
        \and 
            Department of Physics and Astronomy, Clemson University, Kinard Lab of Physics, Clemson, SC 29634-0978, USA
        \and 
            Max-Planck-Institut für extraterrestrische Physik (MPE), Gießenbachstraße 1, 85748 Garching, Germany
        }

   \date{Received ??; accepted ??}

 
  \abstract
    {Protoclusters of galaxies are overdense regions of the Universe characterized by large gas reservoirs. Such environments make them perfect laboratories to investigate galaxy-AGN co-evolution and the growth of SMBHs. Galaxies living in such a dense regions are expected to growth efficiently their SMBH, resulting in a higher incidence of AGN than in the field. Some protoclusters exhibit extended Ly$\alpha$ nebulae in their central region, pinpointing the presence of massive gas reservoirs, but whose main powering mechanism is still debated.}
    {We aim to investigate the AGN population, and AGN enhancement, in three protoclusters at $2.3\lesssim z \lesssim 3.2$, namely the Slug, Fabulous, and J0819, which host enormous ($\approx200-400$ kpc) Ly$\alpha$ nebulae (ELANe). Additionally, we search for the presence of X-ray diffuse emission in the same region of the Ly$\alpha$ nebulae to reveal multi-phase gas in these protoclusters.}
    {We use deep ($t_{\rm exp} \sim 190-270$ ks) \cha observations to identify AGN among the protocluster members and perform X-ray spectral analysis to derive the properties of those sources. We compare the AGN fraction and space density with those observed in other known protoclusters and from the field environment.}
    {Overall, we find 11 X-ray detected AGN in the three protoclusters (2, 5, 4 in the Slug, Fabulous and J0819, respectively). Each structure hosts a central, X-ray powerful (log$(L_{\rm X}/{\rm erg \, s^{-1}}) \sim 45-46$), QSO, while the other X-ray sources are mostly moderately luminous (log$(L_{\rm X}/{\rm erg \, s^{-1}}) \sim 44$) and obscured Compton-Thin AGN. The fraction of AGN in our targets is comparable with estimates for other protoclusters, and significantly higher than what is found for low-redshift clusters. We also find a significant enhancement (2-4 dex) of AGN density with respect to the field and to non active galaxies in the protoclusters. Finally, we find no significant soft X-ray diffuse emissions from the nebulae, thus ruling out gravitational heating as the main powering mechanism of the ELANe.}  
    {}

   \keywords{}

   \titlerunning{AGN in protoclusters}
   \authorrunning{A. Traina et al.}
   \maketitle   
%

\section{Introduction}\label{sec:intro}

Supermassive black holes (SMBHs) play key roles in shaping the properties and evolution of the galaxies in which they are hosted \citep[][]{magorrian1998, ferrarese2002galBH,kormendy2013coev}. This interplay is known as SMBH-galaxy co-evolution \citep[][]{kormendy2013coev}, and is thought to be driven by the phases of strong nuclear activity \citep[][]{aird2012, aird2018,bongiorno2012,aird2015bhard, bongiorno2016,carraro2020}, when the SMBHs trigger as active galactic nuclei (AGN).
\par Additionally, galaxy evolution is also significantly affected by the surrounding environment on scales up to several Mpc.
From a cosmological perspective, cosmic structures evolve hierarchically through the collapse into dense regions, that we observe today as the most massive, gravitationally bound systems, namely galaxy clusters. Within these structures, galaxies are often found to be more evolved than their field analogs \citep[][]{alberts_noble2022}, suggesting an accelerated evolution in their predecessors, that is, galaxy protoclusters \citep[][]{overzier2016,chaing2017}, possibly due to their large gas reservoirs and high merger rates \citep[][]{umehata2019,liu2023}. These condition are thought to foster the SMBHs growth in high-redshift overdense regions \citep[see e.g.,][]{martini2013,assef2015,hennawi2015,marchesi2023,elford2024,vito2024,travascio2025}. In this framework, dedicated studies of the galaxy and AGN populations in protoclusters may unveil key aspects of the connection between galaxies and SMBHs evolution.

Galaxy protoclusters have been typically identified through different probes: high redshift radio galaxies often lie in the core of these structures \citep[][]{pentericci2000,venemans2007,overzier2008,hatch2014,gilli2019}; overdensities of gas-rich sub-millimeter galaxies (SMGs), Ly$\alpha$ emitters (LAEs), Lyman break galaxies (LBGs) \citep[][]{debreuck2004,ouchi2005,dannerbauer2014,chiang2015,oteo2018,umehata2017ncounts, pensabene2024,galbiati2025}
or via spectro-photometric surveys \citep[][]{steidel2000,steidel2005, chiang2014,cucciati2014,toshikawa2016,kashino2023}. 
One of the main pieces of evidence supporting the presence of huge reservoirs of gas in protoclusters is the discovery of diffuse Ly$\alpha$ emission, extending from tens to hundreds of kpc in the intergalactic medium often referred to as the so-called Ly$\alpha$ blobs \citep[][]{geach2009,prescott2008,cantalupo2014,hennawi2015, badescu2017,arrigoni2018,arrigoni2019,umehata2019}. 
Such nebular emission is thought to pinpoint the presence of massive \citep[$\sim 10^{10}$ M$_{\odot}$, see e.g., ][]{prochaska2014}, relatively cold gas inflows infalling from the forming cosmic web towards the center of the potential well. Several physical mechanisms can power this Ly$\alpha$ emission, including recombination radiation after ionization from close ionizing sources \citep[e.g., luminous AGN; e.g.,][]{geach2009,overzier2013,cantalupo2017}, resonant scattering of Ly$\alpha$ photons emitted by a central source \citep[e.g.,][]{borisova2016,costa2022}, shocks and galactic outflows \citep[][]{mori2004} and cooling radiation following collisional excitation during the gravitational collapse of the gas \citep[e.g.,][]{haiman2009}. The availability of gas in protoclusters is thought to fuel fast SMBH accretion in the protocluster galaxy members. This expectation can be tested by studying the AGN population in those structures. X-ray observations are among the best tools to select AGN, being less affected by galaxy dilution and obscuration than other selection methods
\citep[e.g.,][]{brandt2015,padovani2017}.
 In the last decades, deep \cha observations have been carried out with the aim of revealing the AGN population in several protoclusters \citep[see e.g.,][]{geach2009,lehmer2009,digby-north2010,lehmer2013,macuga2019,tozzi2022a,monson2023,vito2020,vito2024,travascio2025}. These studies have generally found an enhancement of SMBH growth in overdense regions with respect to the field environment and low$-z$ clusters \citep[][]{martini2006,bufanda2017}, although with significant scatter. Moreover, X-rays observations can help in shedding light on the origin of the extended Ly$\alpha$ emission in protoclusters. Indeed, the ratio between the X-ray and Ly$\alpha$ luminosities of the diffuse Ly$\alpha$ blob carries useful information on the leading mechanism that powers a nebula \citep[][]{cowie1980,bower2004,geach2009}. 
\par In this work, we present new deep ($t_{exp} \sim 200\,-\,280$ ks) \cha observation towards three fields (namely, the Slug, Fabulous and J0819 protoclusters)  at $z=2.3-3.2$ \citep[][]{cantalupo2014, arrigoni2018, arrigoni2019,chen2021} selected as overdensities of SMGs and LAEs around luminous optically selected AGN. These structures host in their central region some of the largest (200-400 kpc) Ly$\alpha$ nebulae known to date \citep[][]{cantalupo2014,arrigoni2018,arrigoni2019}, which are usually referred to as enormous Ly$\alpha$ nebulae \citep[ELANe;][]{cai2017}.
In this work, we aim at characterizing the population of X-ray detected AGN in these structures from a physical and statistical perspective. We present the properties inferred from the X-ray spectral analysis of the detected sources, and compare the observed overdensity of AGN with the AGN population in the fields, to investigate potential enhancement in SMBH growth in these overdense regions. Finally, to characterize the emission from the Ly$\alpha$ nebula, we take advantage of the \cha sensitivity to investigate the presence of possible diffuse X-ray emission at low energies, spatially corresponding to the ELAN. 
\par The paper is organized as follows. In Section \ref{sec:targets} we describe the targets of the \cha observations, whose analysis and results are presented in Sections \ref{sec:data_analysis} and \ref{sec:results}. In Section \ref{sec:discussion} we discuss the results on the AGN fraction in protoclusters and on the enhancement of SMBHs growth. Finally, we summarize our work in Section \ref{sec:conclusion}. In this work, we assume a \cite{chabrier2003imf} stellar initial mass function (IMF) and adopt a $\Lambda$CDM cosmology with $H_{0} = 67.4$ $\rm km$ $\rm s^{-1}$ $\rm Mpc^{-1}$, $\Omega_{\rm m} = 0.3$, and $\Omega_{\Lambda} = 0.7$ \citep[][]{planck2020}.

\section{Data availability and targets description}\label{sec:targets}

Our sample consists of three protoclusters at $z \sim 2.3 \,-\, 3.2$ surrounding a central quasar (QSO hereafter) embedded in ELANe, with maximum projected extensions of $\sim 150 \,-\,460$ kpc. For these three protoclusters, in addition to the proprietary X-ray observations (PI: F. Vito), we collected catalogs of LAEs, SMGs (detected either by SCUBA-2 or ALMA) and known QSOs from \citet{cantalupo2014,arrigoni2018,arrigoni2019,chen2021,nowotka2022,arrigoni2022,arrigoni2023} and by private communications. We show in Figure \ref{fig:maps} the \cha X-ray images with contours corresponding to the ancillary data available. Throughout the paper, we assume that the galaxies in the prior catalogues are members of the protoclusters, even when the redshift is not spectroscopically confirmed. However, we underline that interlopers may be a source of contamination for the SMGs members \citep[see e.g.,][]{pensabene2024}, even though a small number of field galaxies is expected to fall into the volumes of our protoclusters \citep[$N_{\rm field} \sim 1$, computed using the IR-LFs by][]{Traina2024sfrd}. Regarding possible contaminant on the LAEs survey, since our LAE catalogue (available only for the Slug) has been built from narrow-band data, it could be contaminated by interloping [OII] emitters at $z=0.07$ and CIV emitters at $z=1.58$. At the corresponding redshift, the luminosity limit and survey volume for [OII] and [CIV] correspond to a $3.8 \times 10^{37}$ erg s$^{-1}$, 1.434 cMpc$^3$ and $5.2 \times 10^{40}$ erg s$^{-1}$, 356.4 cMpc$^3$, respectively. By extrapolating the current known luminosity functions for [OII] \citep[e.g.,][]{ciardullo2013} and [CIV] emitters \citep[e.g.,][]{stroe2017} at these redshifts to the much fainter luminosities here probed, we estimate that the contamination should be very low (0.3 and 0.1 galaxies).

In this work, we label sources using the same convention as in the aforementioned catalogs. In this section, we list and discuss briefly the available member catalogs, along with the selection of the protoclusters and their general physical and observational properties. 

\subsection{Slug}\label{subsec:slug_sample}
The Slug ELAN, at $z = 2.2825$, has been discovered by \cite{cantalupo2014}  using custom narrow band filters (NB3985) with the low resolution imaging spectrometer (LRIS) on the Keck I telescope, targeting the radio-quiet QSO UM287. The Ly$\alpha$ nebula is particularly extended ($\sim 460$ kpc) and gas rich \citep[$M_{\rm gas} \sim 10^{12}$ M$_{\odot}$,][]{cantalupo2014}. In addition to UM287, the ELAN embeds an optically-faint, radio-loud QSO. For the Slug protocluster, member catalogs consisting of 45 LAEs and 10 SMGs (seven detected by SCUBA-2 and three through ALMA observations, with continuum detection) have been presented by \cite{cantalupo2014,chen2021} and \cite{nowotka2022}. The total number of unique members is 55, with the SCUBA-2 detected galaxies and LAEs catalog having no members in common, while 2 of the 3 ALMA detected members are also identified as LAEs. Out of the 45 LAEs detected by \cite{cantalupo2014}, 12 have been spectroscopically confirmed to be secure protocluster members (observed with the Binospec on MMT telescope, Chen et al., in preparation). Two ALMA detected galaxies have spectroscopic redshifts based on CO(4-3) line detection, while the other is detected only in continuum. Finally, the SMGs reported in \cite{nowotka2022}, observed with SCUBA-2 as part of the AMUSE$^2$ \citep[A MUltiwavelength Study of ELAN Environments][]{chen2021,arrigoni2022} program, have not been confirmed yet to be secure protocluster members. We also use proprietary LBT data to perform SED fitting for the Slug members candidates, which allow us to obtain informations on the main physical properties of these galaxies (see Appendix \ref{app:seds}).

\subsection{Fabulous}\label{subsec:fabulous_sample}
A similar approach to the one used for the Slug has been used by \cite{arrigoni2018,arrigoni2019} 
targeting Ly$\alpha$ nebulae around $z \sim 3$ QSOs. Using the Multi-Unit Spectroscopic Explorer \citep[MUSE,][]{bacon2010} instrument at VLT, two out of the 61 observed QSOs have been found to be embedded in ELANe, namely the Fabulous and J0819 nebulae. The Fabulous ELAN has a projected size of $\sim 300$ kpc. Using ALMA, SMA, HAWK-I and MUSE, \cite{arrigoni2022,arrigoni2023} discovered a protocluster candidate consisting of 53 sources detected at $850\,\mu$m, 10 of which were also detected at $450\,\mu$m. Moreover, 5 sources, named QSO, QSO2, AGN1, LAE1 and LAE2, have been identified spectroscopically by \citet{arrigoni2018} through their MUSE spectra. 

\subsection{J0819}\label{subsec:j0819_sample} 
The J0819 ELAN is the second nebula discovered by \citet{arrigoni2018,arrigoni2019} and extends for $\sim 150$ kpc. No other optical source in addition to the targeted QSO is embedded in this ELAN. The candidate protocluster surrounding it has been identified by \cite{arrigoni2023} with SCUBA at $450\,\mu$m and $850\,\mu$m, with 5 and 25 sources detected, respectively, for a total of 26 unique sources (25 SMGs and 1 QSO). For a subsample of 16 SMGs (selected as $S/N>4.5$ sources in the SCUBA-2 discovery dataset), shallow observations with ALMA have been conducted (\textcolor{blue}{Y.-J. Wang et al., submitted}), spectroscopically confirming six of these sources as member of the protocluster (the others are not detected in CO(4-3)). However, for the other SMGs a spectroscopic redshift is not available.

\section{Data analysis}\label{sec:data_analysis}

\subsection{X-ray observations and data reduction}\label{subsec:xray_data_reduction}

\begin{table}[]
\centering
\renewcommand{\arraystretch}{1.5}
\caption{Details on the individual \cha observations.}
\begin{threeparttable}
\resizebox{\linewidth}{!}{%
\begin{tabular}{ccccccccc}
\hline \hline
\multicolumn{3}{c}{\textbf{Slug}}                                      & \multicolumn{3}{c}{\textbf{Fabulous}}                                  & \multicolumn{3}{c}{\textbf{J0819}}                       \\ \hline
OBSID                & T$_{\rm exp}$ & \multicolumn{1}{c|}{Date}       & OBSID                & T$_{\rm exp}$ & \multicolumn{1}{c|}{Date}       & OBSID     & T$_{\rm exp}$     & Date                     \\
\multicolumn{1}{l}{} & {[}ks{]}      & \multicolumn{1}{l|}{}           & \multicolumn{1}{l}{} & {[}ks{]}      & \multicolumn{1}{c|}{}           &           & {[}ks{]}          & \multicolumn{1}{l}{}     \\ \hline
26730                & 39.0          & \multicolumn{1}{c|}{2022-12-15} & 26731                & 12.0          & \multicolumn{1}{c|}{2023-03-03} & 26732     & 37.1              & 2024-03-20               \\
27130                & 10.0          & \multicolumn{1}{c|}{2023-10-10} & 27135                & 24.0          & \multicolumn{1}{c|}{2023-11-29} & 27142     & \phantom{0}9.1               & 2024-01-18               \\
27131                & 10.1          & \multicolumn{1}{c|}{2023-10-13} & 27136                & 22.7          & \multicolumn{1}{c|}{2023-12-24} & 27143     & 10.6              & 2024-01-18               \\
27132                & 15.2          & \multicolumn{1}{c|}{2023-10-13} & 27137                & 10.8          & \multicolumn{1}{c|}{2024-04-02} & 27144     & 14.9              & 2023-01-28               \\
27133                & 10.4          & \multicolumn{1}{c|}{2023-09-29} & 27138                & 10.1          & \multicolumn{1}{c|}{2024-02-29} & 27145     & 14.9              & 2024-05-31               \\
27134                & 12.1          & \multicolumn{1}{c|}{2023-10-14} & 27139                & 16.1          & \multicolumn{1}{c|}{2023-03-04} & 27146     & 17.1              & 2024-02-16               \\
28935                & 11.0          & \multicolumn{1}{c|}{2023-10-01} & 27141                & 10.1          & \multicolumn{1}{c|}{2024-02-14} & 27147     & 14.4              & 2023-01-13               \\
28936                & \phantom{0}7.6           & \multicolumn{1}{c|}{2023-10-01} & 27714                & 13.6          & \multicolumn{1}{c|}{2024-03-09} & 27148     & 10.1              & 2024-01-29               \\
28949                & 12.2          & \multicolumn{1}{c|}{2023-10-04} & 27715                & 10.9          & \multicolumn{1}{c|}{2024-03-14} & 27149     & 11.0              & 2024-04-20               \\
28950                & 13.1          & \multicolumn{1}{c|}{2023-10-04} & 29092                & 20.0          & \multicolumn{1}{c|}{2023-11-29} & 27656     & 13.6              & 2023-01-25               \\
28968                & 10.0          & \multicolumn{1}{c|}{2023-10-11} & 29130                & 10.1          & \multicolumn{1}{c|}{2023-12-25} & 27682     & 13.0              & 2023-01-30               \\
28969                & 10.0          & \multicolumn{1}{c|}{2023-10-12} & 29270                & \phantom{0}9.9           & \multicolumn{1}{c|}{2024-02-14} & 29197     & 13.1              & 2024-01-16               \\
28970                & 12.0          & \multicolumn{1}{c|}{2023-10-13} & 29271                & \phantom{0}9.8           & \multicolumn{1}{c|}{2024-02-17} & 29198     & 14.1              & 2024-01-16               \\
28971                & 16.1          & \multicolumn{1}{c|}{2023-10-16} & 29290                & 10.1          & \multicolumn{1}{c|}{2024-02-29} & 29199     & 10.6              & 2024-01-21               \\
\multicolumn{3}{c|}{Total T$_{\rm exp}$ {[}ks{]} = 188.8}              & 29291                & 10.1          & \multicolumn{1}{c|}{2024-03-01} & 29216     & \phantom{0}9.9               & 2024-01-29               \\
                     &               & \multicolumn{1}{c|}{}           & 29312                & 11.6          & \multicolumn{1}{c|}{2024-03-09} & 29217     & 10.1              & 2024-01-29               \\
                     &               & \multicolumn{1}{c|}{}           & 29358                & 18.3          & \multicolumn{1}{c|}{2024-04-08} & 29252     & 10.1              & 2024-02-11               \\
                     &               & \multicolumn{1}{c|}{}           & \multicolumn{3}{c|}{Total T$_{\rm exp}$ {[}ks{]} = 230.2}              & 29253     & 10.1              & 2024-02-11               \\
                     &               & \multicolumn{1}{c|}{}           &                      &               & \multicolumn{1}{c|}{}           & 29272     & 15.1              & 2024-02-16               \\
                     &               & \multicolumn{1}{c|}{}           &                      &               & \multicolumn{1}{c|}{}           & 29344     & 14.7              & 2024-03-28               \\
\multicolumn{3}{c|}{}                                                  & \multicolumn{3}{c|}{}                                                  & \multicolumn{3}{c}{Total T$_{\rm exp}$ {[}ks{]} = 273.6} \\ \hline
                     &               &                                 &                      &               &                                 &           &                   &                         
\end{tabular}}
\begin{tablenotes}
   \small{ \item[*]{Observation ID, exposure times (before background filtering) \\and starting date of each \cha observations.}}  
\end{tablenotes}
\end{threeparttable}
\label{tab:log_chandra_obs}
\end{table}

The three protoclusters (Slug, Fabulous and J0819) were observed with \cha from the end of the 2022 to the mid 2024. Each observation is divided into a number of sub-pointings, identified by their OBSID. The information on the individual observations (OBSID, exposure time and starting date) is summarized in Table \ref{tab:log_chandra_obs}. In this Section, we describe the reduction procedure to obtain the fully calibrated images and the astrometric corrections applied to them. For the analysis, we used the \texttt{ciao\_contrib.runtool} \citep[][version 4.17]{fruscione2006} \texttt{PYTHON} package.
\par We first reprocess the data using the \texttt{chandra\_repro} tool. Then, we filtered each file to remove background flares using \texttt{dmextract} with a 3$\sigma$ threshold. Then, we created relatively astrometrized, merged images. This is done by following this procedure: {\it (i)} a full-band ($0.5 -7$ keV) image, PSF and exposure maps are created for each event file with \texttt{fluximage}; {\it (ii)} we apply \texttt{wavdetect} to the maps using the default significance threshold for detection (\texttt{sigthresh} $=10^{-6}$), to detect sources in the images; {\it (iii)} we match the sources with more than 5 net counts and a PSF size smaller than 5$^{\prime\prime}$ in the different images to compute the offsets with respect to the longest observation (which we use as a reference observation for the relative astrometry correction), by applying \texttt{wcs\_match}; {\it (iv)} finally, we apply these corrections ($\sim 0.5 -0.9^{\prime\prime}$) with wcs\_update, merge all the relatively astrometrized event files and produce images and exposure maps with \texttt{reproject\_obs} and \texttt{flux\_obs}. Finally, we correct the absolute astrometry by repeating the procedure above extracting the sources from the merged datasets and registering their positions on the GAIA DR3 catalog \citep[][]{Gaia2023}. This procedure produces images in the three bands typically used for X-ray studies (i.e., “soft”, “hard” and “full” bands), at $0.5-2$ keV (soft), $2-7$ keV (hard) and $0.5-7$ keV (full).

\subsection{Source detection}\label{subsec:source_catalog}

For each protocluster, we searched for X-ray counterparts of the sources in the prior catalog by running \texttt{wavdetect} on the images ($\sim 16^{\prime} \times 16^{\prime}$, equivalent to $\sim 8 \times 8$ Mpc and $\sim 7.5 \times 7.5$ Mpc at $z \sim 2.3$ and $z\sim 3.2$, respectively). We followed what previously done by \cite{tozzi2022a} and \cite{travascio2025} for the source detection, by setting the values of \texttt{wavdetect.scales} to 1, 1.414, 2, 2.828, 4, 5.657, 8, 11.31 and 16, while the \texttt{wavdetect.sigthresh} parameter was set to $10^{-5}$. We detected 147, 155 and 165 sources in the Slug, Fabulous and J0819 protoclusters, respectively. These sources were matched with the SMGs and LAEs prior catalogs (with a matching radius corresponding to the uncertainty on the position from the prior $\sim 2-4^{\prime \prime}$), resulting into 2, 5 and 4 associations, respectively. Cutouts of the individual sources, in each band, are shown in Figure \ref{fig:cutouts_all}, while their observational properties are reported in Table \ref{tab:observational_quantities}. 
\par For the Slug protocluster, two X-ray associations have been found. UM287 is the bright QSO at the center of the ELAN, while compQSO is the second already known QSO (Section \ref{subsec:slug_sample}). Both of them were detected in the LRIS observations by \citet{cantalupo2014} and were also detected with ALMA by \citet{chen2021}. These two sources are quite bright, with $\sim 500$ and $120$ counts in the full band, respectively. The Fabulous protocluster has a richer AGN content, with 5 X-ray sources associated with its member galaxies. Two of them, QSO\_Fabulous ($\sim 340$ counts) and QSO2 ($\sim 34$ counts), are the central and a nearby QSO, respectively, and were already known as AGN (Section \ref{subsec:fabulous_sample}). The other three matches were not previously known as AGN and are  SMGs detected by SCUBA-2 at $850 \mu$m. Two out of three of these SMGs are detected in the soft, hard and full bands, while the remaining is a hard source without a soft band detection.
In the J0819 protocluster, we detected four X-ray AGN, one of which is the already known central QSO. This source, QSO\_J0819, is very bright both in the soft and in the hard bands, with more than $\sim 1300$ net counts in the full band. One of the other sources is not detected in the hard band, while the other two are detected in both. All of them are SMGs, detected either at $850 \mu$m or $450 \mu$m. J0819-2 has, in addition, a spectroscopic redshift measured by ALMA, confirming it as a protocluster member.

\begin{table*}[]
\centering
\renewcommand{\arraystretch}{1.5}
\caption{Properties of the X-ray detected AGN in the three protoclusters.}
\begin{threeparttable}
\resizebox{\linewidth}{!}{%
\begin{tabular}{cccccccccl}
\hline \hline
\begin{tabular}[c]{@{}c@{}}Source ID\\  (a)\end{tabular} & \begin{tabular}[c]{@{}c@{}}R.A.\\  (b)\end{tabular} & \begin{tabular}[c]{@{}c@{}}DEC.\\ (c)\end{tabular} & \begin{tabular}[c]{@{}c@{}}Prior\\  (d)\end{tabular} & \begin{tabular}[c]{@{}c@{}}spec$-z$\\  (e)\end{tabular} & \begin{tabular}[c]{@{}c@{}}Known AGN ?\\  (f)\end{tabular} &\begin{tabular}[c]{@{}c@{}}$C_{\rm S}$\\  (g)\end{tabular} & \begin{tabular}[c]{@{}c@{}}$C_{\rm H}$\\  (h)\end{tabular} & \begin{tabular}[c]{@{}c@{}}$C_{\rm F}$\\  (i)\end{tabular} & \begin{tabular}[c]{@{}c@{}}HR\\ (l)\end{tabular} \\ \hline 
\multicolumn{9}{c}{\textbf{Slug}}                                                                                                                                                                                        \\ \hline
compQSO                                                       & 0:52:03.24                                          & +1:01:08.60     &     SMG/LAE        &   $2.2825$     &  Yes                   & $59.6_{-7.4}^{+8.1}$                                      & $58.3_{-7.3}^{+7.9}$                                      & $117.9_{-10.5}^{+11.2}$                                   & $-0.05_{-0.27}^{+0.28}$   \\                       
UM287                                                      & 0:52:02.42                                          & +1:01:29.34     &     QSO/ALMA        &  $2.2825$ &     Yes                & $259.6_{-15.7}^{+16.4}$                                   & $233.9_{-14.9}^{+15.6}$                                   & $493.3_{-21.8}^{+22.5}$                                   & $-0.04_{-0.27}^{+0.28}$                          \\ \hline 
\multicolumn{9}{c}{\textbf{Fabulous}}                                                                                                                                                                                                                                                                                                                                                                                                                                                                                                                                                                                                                                                                                                                                                                                 \\ \hline
QSO\_Fabulous                                     & 10:20:10.00                                                      & +10:40:02.70     &     QSO/SMG        & $3.164$        &     Yes                          & $158.6_{-12.2}^{+12.9}$                                                             & $181.8_{-13.1}^{+13.8}$                                                              & $340.4_{-18.1}^{+18.7}$                                                             & $0.06_{-0.28}^{+0.28}$ \\                                                   
Fab-21                                            & 10:19:58.90                                                       & +10:44:54.00    &     SMG        &           &     No                           & $3.6_{-1.7}^{+2.4}$                                                              & $6.3_{-2.3}^{+3.0}$                                                              &  $9.9_{-3.0}^{+3.7}$                                                            & $0.04_{-0.28}^{+0.27}$   \\                                                
Fab-33                                            & 10:20:19.50                                                       & +10:39:04.00    &     SMG        &          &      No                         & $<5.2$                                                              & $8.6_{-2.7}^{+3.4}$                                                              & $10.4_{-3.0}^{+3.7}$                                                              & $>0.74$    \\                                                
Fab-41                                            & 10:20:06.20                                                      & +10:38:30.00     &     SMG        &          &      No                         & $4.4_{-1.9}^{+2.6}$                                                              & $10.1_{-3.0}^{+3.7}$                                                              & $15.6_{-3.8}^{+4.5}$                                                             & $0.16_{-0.28}^{+0.26}$   \\                                                 
QSO2                                                       & 10:20:09.56                                                      & +10:40:05.30     &     QSO        & $3.158$           &     Yes                            & $16.7_{-3.9}^{+4.5}$                                                              & $16.8_{-3.9}^{+4.6}$                                                              & $33.6_{-5.6}^{+6.2}$                                                           & $-0.01_{-0.27}^{+0.28}$         \\ \hline                                            
\multicolumn{9}{c}{\textbf{J0819}}                                                                                                                                                                                                                                                                                                                                                                                                                                                                                                                                                                                                                                                                                                                                                                                    \\ \hline
QSO\_J0819                                                 & 08:19:40.60                                                       & +8:23:58.00   &    QSO        & 3.213            &     Yes                           & $671_{-25.9}^{+25.9}$                                                             & $695_{-26.4}^{+26.4}$                                                              & $1366_{-36.9}^{+36.9}$                                                             & $0.06_{-0.27}^{+0.27}$  \\                                                  
J0819-24                                      & 08:19:21.00                                                      & +8:20:56.00    &    SMG        &            &      No                         & $4.1_{-1.9}^{+2.6}$                                                             & $19.2_{-4.3}^{+4.9}$                                                             & $23.4_{-4.8}^{+5.4}$                                                             & $0.42_{-0.26}^{+0.21}$    \\                                                
J0819-16                                      & 08:19:43.87                                                      & +8:23:51.74    &    SMG        &            &      No                         & $3.5_{-1.7}^{+2.4}$                                                             & $<4.8$                                                             & $4.9_{-2.2}^{+2.8}$                                                            & $<0.25$   \\ 
J0819-2                                       & 08:19:37.23                                                      & +8:23:05.76    &    SMG        &  3.208          &      No                             & $13.5_{-3.4}^{+4.1}$                                                             & $20.3_{-4.3}^{+4.9}$                                                             & $33.9_{-5.6}^{+6.2}$                                                             & $0.12_{-0.29}^{+0.27}$                 \\ \hline \hline \\                                    
\end{tabular}}
\begin{tablenotes}
   \small{ \item[*]{Name (a), coordinates (b, c), type of prior source (d), spectroscopic redshift (e), previous information on the AGN nature of the source (f),\\ net counts in the soft (g), hard (h) and full (i) bands (with $1 \sigma$ errors) and the hardness ratios (l) for the X-ray AGN in the three protoclusters.}}  
\end{tablenotes}
\end{threeparttable}
\label{tab:observational_quantities}
\end{table*}

\section{Results}\label{sec:results}

\subsection{X-ray spectral properties}\label{subsec:xray_properties}

\begin{figure}[]
{\includegraphics[width=.5\textwidth]{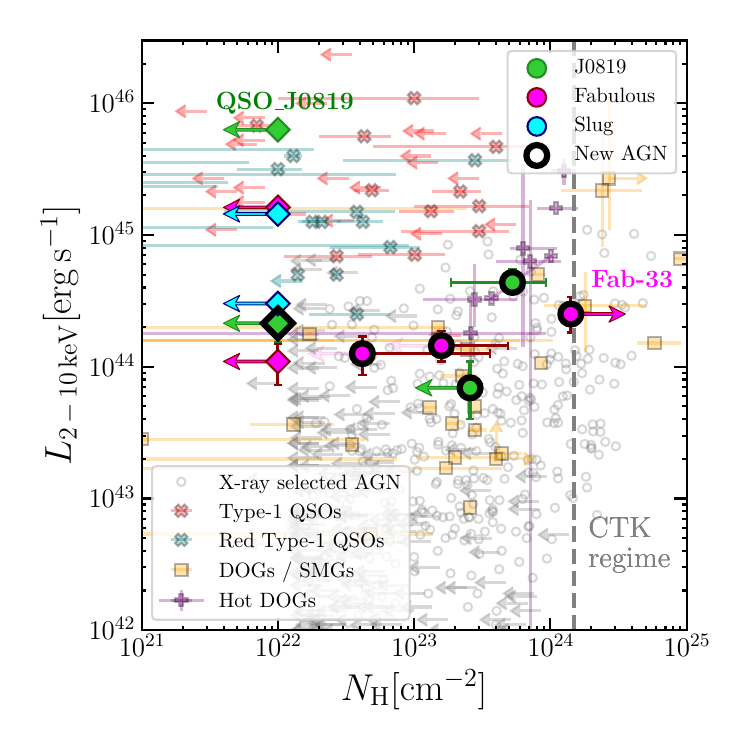}}
\caption{Intrinsic $2-10$ keV luminosity against column density. The protocluster members reported here are are shown in cyan, magenta and green. Diamonds indicate sources whose properties were derived via spectral fitting, while circles are for sources with properties derived via HR. Other classes of AGN detected in the X-ray are shown in different colors \citep[][]{urrutia2005,just2007,wang2013,banerji2014,stern2014,assef2016,corral2016,martocchia2017,mountrichas2017,ricci2017,goulding2018,vito2018b,zappacosta2018,li2019chandra,lansbury2020,zou2020}. Newly discovered AGN (i.e., not known as such prior to this work) are highlighted with thick black circles.}
\label{fig:LX_NH}
\end{figure}

In this section, we describe the properties of the X-ray detected AGN in the protoclusters. For the five AGN with a total number of counts $C_{\rm F} > 30$ in the $0.5-7$ keV band, we performed a spectral fitting with \texttt{pyXspec} \citep[][]{gordon2021pyxspec}.
We extracted source spectra, Ancillary Response Files (ARFs) and Response Matrix Files (RMFs), for each OBSID with \texttt{specextract} tool and then combined using \texttt{combine\_spectra} tool in the CIAO \texttt{PYTHON} package. Spectra have been extracted from regions with radius of 2$^{\prime\prime}$ for sources near the center of the image ($\theta <$ 5$^{\prime}$) or using a region with radius of 3$^{\prime\prime}$ for sources in the outer regions ($5^{\prime} <\theta < 15^{\prime\prime}$), corresponding to an encircled energy fraction EEF $\approx 90\%$. ARF files have been corrected for the lost PSF fraction. Background regions have been selected to be near the source but uncontaminated by its emission. We show the extracted spectra in Figure \ref{fig:spectra_all}, along with the best fitting simple powerlaw model (e.g., phabs $\times$ zpha $\times$ po, where phabs is the Galactic absorption column density\footnote{\hyperlink{https://heasarc.gsfc.nasa.gov/cgi-bin/Tools/w3nh/w3nh.pl}{https://heasarc.gsfc.nasa.gov/cgi-bin/Tools/w3nh/w3nh.pl}}). Results from the spectral analysis are reported in Table \ref{tab:spectral_properties}. The main parameters derived are photon indices ($\Gamma$), intrinsic column densities along the line of sight ($N_{\rm H}$), observed $0.5 -7$ keV fluxes and intrinsic, unabsorbed, X-ray luminosities between 2 and 10 keV. For the remaining six sources with $C_{\rm F} < 30$, we derived the main X-ray properties via their hardness ratios $HR = (H-S)/(H+S)$, where $H$ is the number of hard counts and $S$ is the number of soft counts. We estimated an “effective” photon index and column density \citep[assuming an intrinsic photon index $\Gamma = 1.9$, see e.g.,][]{vito2024} using the $HR$. In addition, since we did not perform a spectral fit on these sources, we computed the observed flux and intrinsic luminosity by assuming $\Gamma = 1.9$ and the effective $N_{\rm H}$ obtained from the $HR$ analysis. The $HR$ values and the inferred parameters are reported in Table \ref{tab:observational_quantities}. We note that for sources with $>30$ counts the parameters obtained via spectral and HR analysis are consistent within the uncertainties. Hereafter, for the sources with more than 30 counts we use the results from the spectral fitting, whereas for fainter sources we use the results obtained with the $HR$. 
Below we summarize the X-ray spectral properties for the three protoclusters. 
\\
\\
-- Slug. In the Slug protocluster, UM287 and compQSO are luminous, unabsorbed AGN, with $\Gamma \sim 2$ and $\sim 1.6$. Their spectra can be fitted with an unabsorbed powerlaw, without requiring the addition of an absorption component. UM287, like the central QSOs of the other protoclusters, is a powerful source, with $L_{\rm 2-10\,keV} \sim 1.5 \times 10^{45}$ erg s$^{-1}$, corresponding to a bolometric luminosity $L_{\rm BOL} \approx 5 \times 10^{46}$ erg s$^{-1}$ \citep[assuming the bolometric correction by][]{duras2020}.
\\
\\
-- Fabulous. QSO\_Fabulous (i.e., the central QSO) is unabsorbed, with an intrinsic photon index $\Gamma \sim 1.8$, and very bright, similarly to the QSO at the center of the Slug ELAN, with $L_{\rm 2-10\,keV} \sim 1.6 \times 10^{45}$ erg s$^{-1}$ (i.e., $L_{\rm BOL} \approx 5 \times 10^{46}$ erg s$^{-1}$). For QSO2 we fitted a powerlaw with a fixed photon index ($\Gamma \sim 1.9$), leading to similar results with respect to those obtained from the $HR$. Both Fab-21 and Fab-41 are characterized by Compton-Thin column densities and $L_{\rm 2-10\,keV} \sim 1.3 - 2 \times 10^{44}$ erg s$^{-1}$. Fab-33 is instead significantly obscured, being in the Compton-Thick regime (log$N_{\rm H}$ [cm$^{-2}$] $> 24$). Given its large column density, we derived its flux and luminosity by considering a \texttt{MyTORUS} model\footnote{The \texttt{MyTORUS} code is a physically motivated model, widely used in the X-ray spectral analysis of Compton-Thick sources, based on three tables for different components of the spectrum. These components are the zeroth-order emission component, the scattered continuum and the iron line emission component.} \citep[in its coupled configuration, where the column density and the inclination angle of the three components are tied together,][]{murphy2009} in addition of using a classic absorbed powerlaw, and found consistent values. Among the sources previously known in the protocluster, the active galaxy AGN1 \citep[][]{arrigoni2018} is not detected in our X-ray map. This AGN is classified by \cite{arrigoni2018} as a type-2 AGN from its narrow-line spectrum, making it a potentially obscured (and weak) source and, for this reason, being undetected in our data. For this source, we estimated an upper limit on its observed flux and intrinsic luminosity, assuming an intrinsic photon index $\Gamma = 1.9$. We obtained $F_{\rm 0.5-7 \, keV} < 8 \times 10^{-16}$ erg cm$^{-2}$ s$^{-1}$ and corresponding intrinsic luminosity $L_{\rm 2-10 \, keV} < 7.2 \times 10^{43}$ erg s$^{-1}$.
\\
\\
-- J0819. Finally, J0819 hosts one bright, unabsorbed, AGN (QSO\_J0819, at the center). It is the brightest source in our sample of X-ray detected protocluster members and among the most X-ray luminous QSOs known at all cosmic times, with log$L_{\rm 2-10 \, keV}$ [erg s$^{-1}$] $\sim 45.8$ ($L_{\rm BOL} \approx 3.6 \times 10^{47}$ erg s$^{-1}$). J0819-24 and J0819-16 have quite flat effective photon indices and their column densities are in the Compton-Thin regime.
\\
\\
In Figure \ref{fig:LX_NH} we compare the properties ($N_{\rm H}$ and $L_{\rm 2-10 \, keV}$) of our sample with those of other populations of AGN (i.e., Type-1 QSOs, Red Type-1 QSOs, DOGs/SMGs, HotDOGs). The bright AGN/QSOs in our sample are all unabsorbed. The other sources have properties similar to the typical $N_{\rm H}$ and $L_{\rm X}$ of AGN in protoclusters and X-ray selected AGN. We also highlight newly discovered AGN in our sample, showing that X-ray observations are required to unveil part of the hidden population of AGN in protoclusters.
\par For each protocluster, we performed a stacking analysis in the X-ray soft, hard and full bands, of the sources which were not detected in the X-ray maps but were part of the prior catalogs (SMGs and LAEs). In addition, we also stacked SMGs and LAEs separately. No significant emission was found for any of the protoclusters as a result of this stacking analysis, which led to effective exposure times, in the full band, of $\sim 9.5$, $10$ and $6$ Ms.
\par For the optically selected QSOs in the three protoclusters, we also report in Table \ref{tab:spectral_properties} the quantity $\alpha_{\rm ox}=0.38\times\mathrm{log}\left(\frac{L_{2\mathrm{keV}}}{L_{2500\mathrm{\AA}}}\right)$, where $L_{2\mathrm{keV}}$ and $L_{2500\mathrm{\AA}}$ are the mono-chromatic luminosities at rest-frame 2 keV and 2500 $\AA$ \citep[e.g.,][]{brandt2015}. This parameter is widely used to measure the relative emission of the hot corona and accretion disk as a proxy for the accretion physics \citep[e.g.,][]{lusso2017}. A well-known anti-correlation exists between $\alpha_{\rm ox}$ and the UV luminosity of QSOs, implying that the contribution of the hot corona to the total QSO energy output decreases at high luminosities \citep[e.g.,][]{just2007}. For the two QSOs in the Slug nebula, we obtained the $L_{2500\mathrm{\AA}}$ from the observed i-band magnitude from our LBT/LBC observations. For the two QSOs in the Fabulous protocluster, we measured the 1450 $\mathrm{\AA}$ flux from the spectra presented by \cite{arrigoni2018}, and for the optically selected QSO in the J0819 protocluster, we used the magnitude at 1450 $\mathrm{\AA}$ tabulated in \cite{arrigoni2018}. In all cases, we assumed a typical QSO UV continuum $F_\nu\propto\nu^{-\alpha}$ with $\alpha=0.5$ to obtain the luminosity at rest-frame 2500  $\mathrm{\AA}$. The $\alpha_{\rm ox}$ values that we found are in broad agreement with the $\alpha_{\rm ox}$-$L_{\mathrm{UV}}$ anti-correlations published in the literature \citep[e.g.,][]{just2007,lusso2016}. The slightly flat $\alpha_{\rm ox}$ value of compQSO in the Slug protocluster is consistent with its radio-loud nature \citep[][]{cantalupo2014}.

\begin{table}[]
\centering
\renewcommand{\arraystretch}{1.5}
\caption{Properties of the X-ray detected AGN in the three protoclusters with number of counts $> 30$, derived from spectral analysis.}
\begin{threeparttable}
\resizebox{\linewidth}{!}{%
\begin{tabular}{ccccclc}
\hline \hline
\begin{tabular}[c]{@{}c@{}}Source ID\\ \\ (a)\end{tabular} & \begin{tabular}[c]{@{}c@{}}$\Gamma$\\ \\ (b)\end{tabular}  & \begin{tabular}[c]{@{}c@{}}$F_{\rm 0.5-7 \, keV}$\\ $10^{-15}$ erg cm$^{-2}$ s$^{-1}$\\ (c)\end{tabular} & \multicolumn{1}{c}{\begin{tabular}[c]{@{}c@{}}log$(L_{\rm 2-10 \, keV})$\\ erg s$^{-1}$\\ (d)\end{tabular}} & \multicolumn{1}{c}{\begin{tabular}[c]{@{}c@{}}log$(L_{\rm 2500 \, \mathrm{\AA}})$\\ erg Hz$^{-1}$ s$^{-1}$\\ (e)\end{tabular}} & \multicolumn{1}{c}{\begin{tabular}[c]{@{}c@{}}$\alpha_{\rm ox}$\\ \\ (f)\end{tabular}} \\ \hline
\multicolumn{6}{c}{\textbf{Slug}}                                                                                                                                                                                                                                                                                                                                                                                                                                                                                                                                                                                                                                                                                                                                                                                     \\ \hline
compQSO                                                       & $1.63_{-0.39}^{+0.40}$                                                                      & $13.4_{-2.5}^{+1.9}$                                                                                                     & $44.48_{-0.18}^{+0.06}$                                                                                                     & $30.0$                                                                                                     & $-1.27_{-0.02}^{+0.02}$                                                                                                                         \\
UM287                                                      & $2.01_{-0.16}^{+0.17}$                                                                    & $55.5_{-3.8}^{+4.4}$                                                                                                     & $45.16_{-0.04}^{+0.03}$                                                                                                     & $31.8$                                                                                                     & $-1.74_{-0.02}^{+0.01}$                                                                                                                         \\ \hline
\multicolumn{6}{c}{\textbf{Fabulous}}                                                                                                                                                                                                                                                                                                                                                                                                                                                                                                                                                                                                                                                                                              \\ \hline
QSO\_Fabulous                                     & $1.79_{-0.21}^{+0.22}$                                                                      &  $32.6_{-3.0}^{+3.6}$                                                                                                    & $45.21_{-0.04}^{+0.05}$                                                                                                     & $32.0$                                                                                                     & $-1.81_{-0.01}^{+0.01}$                                                                                                                         \\
QSO2                                                       & $1.9^f$                                                                       &  $2.5_{-1.3}^{+0.9}$                                                                                                     & $44.29_{-0.18}^{+0.13}$                                                                                                     & $29.4$                                                                                                     & $-1.16_{-0.03}^{+0.03}$                                                                                                                          \\ \hline
\multicolumn{6}{c}{\textbf{J0819}}                                                                                                                                                                                                                                                                                                                                                                                                                                                                                                                                                                                                                                                                                               \\ \hline
QSO\_J0819                                                 & $1.89_{-0.09}^{+0.11}$                                                                        & $115.2_{-4.4}^{+4.9}$                                                                                                     & $45.80_{-0.02}^{+0.02}$                                                                                                     & $31.5$                                                                                                     & $-1.63_{-0.02}^{+0.01}$                                                                                                                         \\ 
J0819-2                                       & $1.9^{f}$                                                                       &  $5.29_{-1.62}^{+1.48}$                                                                                                     & $44.47_{-0.18}^{+0.13}$                                                                                                                                                                                                                                                   \\ \hline \hline \\                                                                             
\end{tabular}}
\begin{tablenotes}
   \small{ \item[*]{Name (a), photon indices (b), $0.5-7$ keV observed fluxes (c), X-ray\\ luminosities (d), 2500$\AA$ luminosities (e) and $\alpha_{\rm ox}$ (f) for the X-ray AGN \\ in the three protoclusters. Values labeled with $^f$ have been fixed.}}  
\end{tablenotes}
\end{threeparttable}
\label{tab:spectral_properties}
\end{table}

\begin{table}[]
\centering
\renewcommand{\arraystretch}{1.5}
\caption{Properties of the X-ray detected AGN in the three protoclusters derived using their $HR$.}
\begin{threeparttable}
\resizebox{\linewidth}{!}{%
\begin{tabular}{ccccl}
\hline \hline
\begin{tabular}[c]{@{}c@{}}Source ID\\ \\ (a)\end{tabular} & \begin{tabular}[c]{@{}c@{}}$\Gamma_{\rm eff}$\\ \\ (b)\end{tabular} & \begin{tabular}[c]{@{}c@{}}log($N_{\rm H}$ / {\rm [cm$^{-2}$]})\\ \\ (c)\end{tabular} & \begin{tabular}[c]{@{}c@{}}$F_{\rm 0.5-7 \, keV}$\\ $10^{-15}$ erg cm$^{-2}$ s$^{-1}$\\ (d)\end{tabular} & \multicolumn{1}{c}{\begin{tabular}[c]{@{}c@{}}log$(L_{\rm 2-10 \, keV})$\\ erg s$^{-1}$\\ (e)\end{tabular}} \\ \hline
\multicolumn{5}{c}{\textbf{Slug}}                                                                                                                                                                                                                                                                                                                                                                                                                                                                                                                                                                                                                                                                                                                                                                                     \\ \hline
compQSO                                                       & $1.89_{-0.67}^{+0.67}$                                                                    & $<23.63$                                                                                & $15.5_{-1.4}^{+1.5}$                                                                                                     & $44.58_{-0.04}^{+0.04}$                                                                                                                         \\
UM287                                                      & $2.00_{-0.70}^{+0.67}$                                                                   & $<23.57$                                                                               & $59.2_{-2.6}^{+2.7}$                                                                                                     & $45.19_{-0.03}^{+0.02}$                                                                                                                         \\ \hline 
\multicolumn{5}{c}{\textbf{Fabulous}}                                                                                                                                                                                                                                                                                                                                                                                                                                                                                                                                                                                                                                                                                                                                                                                 \\ \hline
QSO\_Fabulous                                     & $1.77_{-0.68}^{+0.68}$                                                                    & $<23.85$                                                                                &  $32.5_{-1.7}^{+1.8}$                                                                                                    & $45.21_{-0.02}^{+0.02}$                                                                                                                         \\
Fab-21                                            & $1.79_{-0.68}^{+0.71}$                                                                    & $22.62_{-*}^{+0.94}$                                                                                &  $2.0_{-0.6}^{+0.7}$                                                                                                    &                      $44.11_{-0.16}^{+0.13}$                                                                                                     \\
Fab-33                                            & $<0.29$                                                                    & $>24.15$                                                                                 & $1.5_{-0.5}^{+0.4}$                                                                                                     & $44.40_{-0.14}^{+0.13}$                                                                                                                         \\
Fab-41                                            & $1.53_{-0.64}^{+0.67}$                                                                    & $23.20_{-*}^{+0.49}$                                                                                &  $1.8_{-0.4}^{+0.5}$                                                                                                    &   $44.19_{-0.12}^{+0.11}$                                                                                                                        \\
QSO2                                                       & $1.94_{-0.69}^{+0.67}$                                                                    & $<23.80$                                                                                 &  $3.6_{-0.6}^{+0.7}$                                                                                                     & $44.29_{-0.07}^{+0.08}$                                                                                                                          \\ \hline
\multicolumn{5}{c}{\textbf{J0819}}                                                                                                                                                                                                                                                                                                                                                                                                                                                                                                                                                                                                                                                                                                                                                                                   \\ \hline
QSO\_J0819                                                 & $1.77_{-0.66}^{+0.66}$                                                                    & $<23.86$                                                                                  & $106.1_{-1.7}^{+1.6}$                                                                                                     & $45.73_{-0.03}^{+0.03}$                                                                                                                         \\
J0819-24                                      & $0.82_{-0.65}^{+0.66}$                                                                    & $23.72_{-0.45}^{+0.25}$                                                                                & $3.8_{-0.8}^{+0.9}$                                                                                                     & $44.64_{-0.09}^{+0.10}$                                                                                                                         \\
J0819-16                                      & $>1.33$                                                                    & $<23.41$                                                                                  & $0.7_{-0.3}^{+0.4}$                                                                                                    & $43.84_{-0.24}^{+0.20}$     \\ 
J0819-2                                       & $1.60_{-0.69}^{+0.67}$                                                                    & $<23.93$                                                                                 &  $7.47_{-1.22}^{+1.37}$                                                                                                     & $44.53_{-0.08}^{+0.09}$                                           \\ \hline \hline \\                                                                             
\end{tabular}}
\begin{tablenotes}
   \small{ \item[*]{Effective photon indices (b), absorption column density (c), $0.5-7$ keV\\ observed fluxes (d) and $2-10$ keV intrinsic luminosities (e) for the X-ray\\ AGN in the three protoclusters. Observed flux and X-ray luminosity for\\ Fab-33 are obtained using \texttt{MyTORUS}.}}  
\end{tablenotes}
\end{threeparttable}
\label{tab:hr_properties}
\end{table}

\subsection{X-ray diffuse emission}\label{subsec:xray_diffuse}

\begin{figure*}[]
\centering
{\includegraphics[width=1.\textwidth]{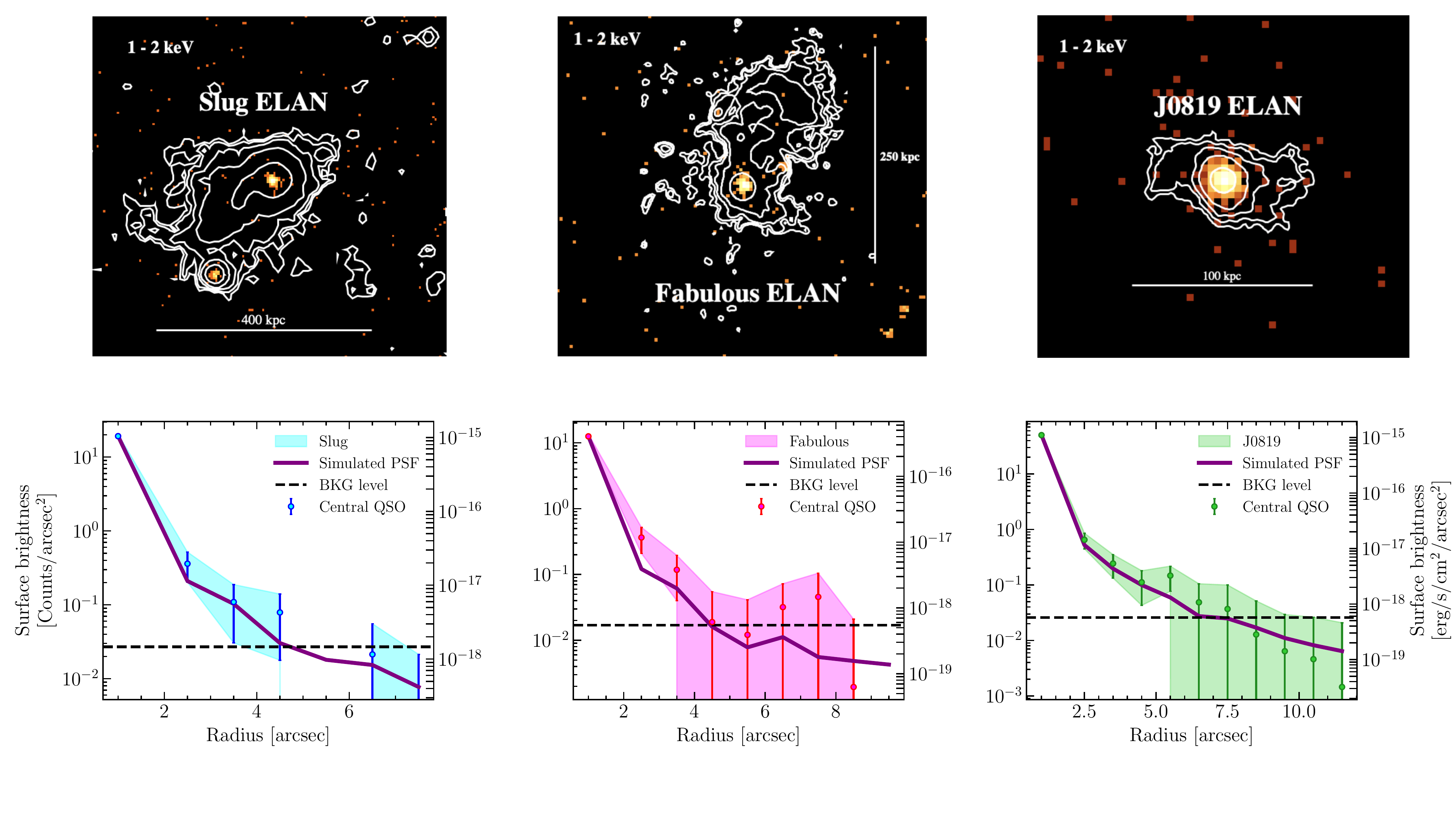}}
\caption{$1-2$ keV maps of the central QSOs of the Slug, Fabulous and J0819 protoclusters (upper panels) with Ly$\alpha$ contours (white) and the background subtracted differential radial profiles (bottom panels) for the emission surrounding the Slug (cyan), Fabulous (magenta) and J0819 (green) central QSOs, with respect to the PSF prediction (purple line). The secondary y-axis is based on the \texttt{apec} model with a fixed temperature of 2 keV.}
\label{fig:ELAN_radial_profiles}
\end{figure*}

The gaseous halo surrounding quasar host galaxies is expected to be multiphase, with at least a cool ($10^4$ K) and a warm-hot ($10^{5-7}$ K) gas components \citep[e.g.,][]{costa2022,obreja2024}. While the cool gas is now routinely traced with Ly$\alpha$ observations, there is little to no information on the warm-hot gas phase at these redshifts. Acquiring information on this component is key in assessing the total baryon budget in the halo, and ultimately to constrain the strength of AGN feedback \citep[e.g.,][]{parimbelli2023}. X-ray emission can in principle probe the hot gas component, but its levels are expected to be low \citep[e.g.,][]{bertone2010}. Additionally, comparing the luminosity of the Ly$\alpha$ halo with that of the co-spatial X-ray emission can help us in disentangling the different origins for this radiation (Section \ref{elane_powering}). The potential diffuse X-ray emission is expected to peak in the soft part of the spectrum, therefore we investigate the presence of extended emission surrounding the central QSOs in the observed $1-2$ keV band. The use of a narrow energy range alleviates the contamination due to the dominant QSO X-ray emission. Figure \ref{fig:ELAN_radial_profiles} shows the X-ray emission in the $1-2$ keV range around the central QSOs of the Slug, Fabulous and J0819. The QSOs clearly correspond to the peak of the Ly$\alpha$ emission from the nebula. Upon visual inspection, QSO\_Fabulous shows a possible excess of counts in the N-S direction (aligned with the nebula) and QSO\_J0819 has a counts excess circularly shaped. UM287 instead does not show any evidence for potential extended emission. 
\par To statistically investigate the nature of this emission, we performed simulations to reconstruct (via ray backtracing) the actual PSF of the observation at the position of the QSO, following a similar approach to \cite{tozzi2022b} \citep[but see also][for a similar approach]{fabbiano2017,traina2021}. We obtained the PSF image with the following procedure. With the \texttt{CHART} online tool \citep[][]{carter2003}, we simulated 10 different PSFs for each OBSID. Then, we used \texttt{MARX} and \texttt{simulate\_psf} to obtain an event file of each simulated image and, eventually, merging each OBSID image to obtain the final merged simulated image of the PSF. Finally, we divided the emission by 10 to get a number of counts compatible with the actual emission. Once the simulated PSF images are produced, it is possible to compare the radial profiles obtained for the observed and simulated images. This comparison allows us to understand whether or not the hint of extended emission around the QSO has a statistical significance or if it can just be ascribed to the PSF wings. Figure \ref{fig:ELAN_radial_profiles} shows this comparison for the three QSOs. In all cases, the X-ray radial profiles of the emission around the QSOs are consistent with the simulated PSF expectations within the errors. We thus conclude that our observations do not detect statistically significant extended X-ray emission at the positions of the ELANe.

\section{Discussion}\label{sec:discussion}

\subsection{AGN fraction}\label{subsec:agn_fraction}

\begin{table}[]
\centering
\renewcommand{\arraystretch}{1.5}
\caption{Fractions of X-ray AGN to SMGs and LAEs.}
\begin{threeparttable}
\begin{tabular}{ccclcl}
\hline
Protocluster & $z$ & \multicolumn{1}{c}{$f_{\rm AGN}^{\rm SMG}$}& \multicolumn{1}{c}{$f_{\rm AGN}^{\rm LAE}$}\\
\hline
Slug & 2.2825 &$0.20_{-0.13}^{+0.20}$& $0.04_{-0.03}^{+0.06}$ ($0.17_{-0.11}^{+0.18}$)\\
Fabulous & 3.1695 &$0.08_{-0.04}^{+0.05}$& \multicolumn{1}{c}{$-$}\\
J0819 & 3.205 & $0.15_{-0.07}^{+0.10}$ ($0.33_{-0.21}^{+0.28}$)& \multicolumn{1}{c}{$-$}\\
\hline
\\

\end{tabular}
\begin{tablenotes}
   \small{ \item[*]{Name of the protocluster, redshift, fraction of X-ray AGN-to-SMG and fraction of X-ray AGN-to-LAEs. The fraction in parenthesis refers to spectroscopically confirmed objects. Errors on the fractions are computed by applying binomial proportion.}}  
\end{tablenotes}
\end{threeparttable}
\label{tab:agn_fractions}
\end{table}

\begin{figure}[]
\centering
\includegraphics[width=.5\textwidth]{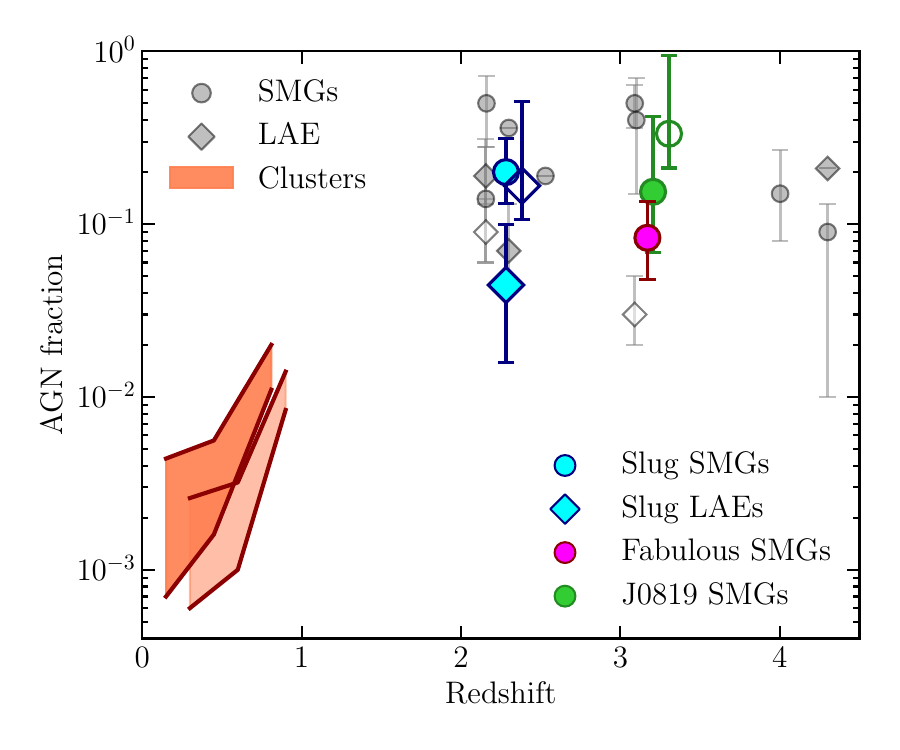}
 \caption{Fractions of X-ray selected AGN with respect to different population of galaxies (i.e., SMGs, LAEs) in the Slug, Fabulous and J0819 protoclusters, compared to other protoclusters \citep[][]{lehmer2009,digby-north2010,smail2014,chen2016,macuga2019,umehata2019,vito2020,polletta2021,tozzi2022a,monson2023,perez-martinez2023,vito2024,travascio2025} and the X-ray AGN fractions in local clusters \citep[][]{martini2006, bufanda2017}. Empty blue and green points are the fractions computed for spectroscopically confirmed members (see Table \ref{tab:agn_fractions}).}
 \label{fig:agn_fractions}
\end{figure}

Several works in the recent years have pointed out an enhancement of the fraction of AGN in protocluster environments with respect to fileds \citep[e.g.,][]{tozzi2022a,vito2024,travascio2025}. We studied the number of AGN in our sample compared to the number of prior sources in different parent samples of SMGs for all protoclusters and LAEs for the Slug, for which a coverage of LAEs is available for the whole field. Uncertainties on the fractions are computed using binomial proportion intervals \citep[][]{brown2001}. The fraction of AGN among SMGs in the Slug is very high ($\sim 20\%$), since each of the three AGN are detected either with ALMA (continuum) or with SCUBA-2. The Fabulous and J0819 have a similar AGN fraction ($\sim 8-12 \%$). For the Slug, we were also able to compute the AGN fraction among LAEs. For the whole sample of LAEs this fraction is quite low ($\sim 0.04$), but higher if we consider just the spectroscopically confirmed LAEs (which correspond to the brightest subsample of LAEs detected with the MMT, $f_{\rm AGN}^{\rm LAE} \sim 0.17$). These results are summarized in Table \ref{tab:agn_fractions} and shown in Figure \ref{fig:agn_fractions}. The fractions we derived agree well with estimates from other works \citep[see][for a compilation]{vito2024}. The Fabulous and J0819 have indeed a fraction of AGN-to-SMGs similar to what is found for protoclusters at $z \sim 3-4.5$, while the higher fraction found for the Slug protocluster is compatible with a slightly higher fraction expected at $z \sim 2-2.5$, possibly linked to an enhancement of the accretion activity at cosmic noon. This result is in agreement with the high fraction of AGN in SMGs (i.e., $f_{\rm AGN} \sim 0.14$) found in the Spiderweb protocluster at similar redshift. Nevertheless, we should stress again that the SMGs in the prior sample \citep[i.e.,][]{nowotka2022} are not spectroscopically confirmed, leading to possibly higher AGN fractions. At higher redshift, the SSA22 protocluster at $z \sim 3.2 $ has the largest fraction of AGN known so far ($f_{\rm AGN} \sim 0.5$). The AGN-to-LAEs fraction is also similar to a lower fraction found in other protoclusters, which is thus consistent with LAEs being expected to reside in lower mass halos than SMGs. 
These results show that the fraction of AGN in the ELAN candidate protocluster is consistent with that of other protocluster, and significantly higher than for low$-z$ clusters.

\subsection{X-ray AGN luminosity function}\label{subsec:xray_agn_lf}
As we highlighted in Section \ref{sec:intro}, protoclusters have been found to foster SMBHs growth thanks to the large availability of gas. This enhancement translates into the XLF normalization and shape \citep[][]{tozzi2022a,travascio2025,vito2024}. Indeed, previous works estimated the volume density (log$\Phi$ [Mpc$^{-3}$ dex$^{-1}$]) of AGN in protoclusters to be from 1 \citep[$z \sim 2$,][]{tozzi2022a} to 4 \citep[$z \sim 4$,][]{vito2024} dex larger than in the field. Further investigations of the XLF at various redshifts are of key importance to understand on how, and how quickly, BH build up their masses in protoclusters at different cosmic epochs. In this section, we derived the volume density of the AGN in our sample, at $z \sim 2.2$ (from the Slug protocluster) and at $z \sim 3.2$ (from the Fabulous and J0819 AGN population) and compare these results with estimates obtained for other protoclusters at similar redshifts and the field environment. 
\par For the estimation of the volume of our protoclusters we considered two possible approaches: {\it i)} we choose as the comoving volume of each protocluster the same used by \citet{travascio2025} for MQN01 (which has among the best multiwavelength and spectroscopical coverage) to make a coherent comparison with their work, that is $V = 16$ cMpc$^2 \, \times \, 2000$ km s$^{-1}$ (we will refer to this volume as the “observed” volume). The comoving area (16 cMpc$^2$) corresponds to the area of the MUSE field-of-view for MQN01 ($\sim 2^{\prime} \times 2^{\prime}$), and a size along the line of sight of $\pm 1000$ km s$^{-1}$. For each protocluster, we converted the physical area into an angular size depending on the redshift of each system and considered only the sources falling into the 16 cMpc$^2 \, \times \, 2000$ km s$^{-1}$ cube. Using this definition for the volume allow us to coherently compare with a few works that focused on the inner regions of protoclusters; 
{\it ii)} Secondly, we use as the radius of the protoclusters, the predicted comoving radius at the relevant redshift in the simulation of \cite{muldrew2015}. The first, observed volumes correspond to 488, 441 and 440 cMpc$^3$ for the Slug, Fabulous and J0819, respectively. The second estimates are obtained from the Millennium Simulation \citep[][]{springel2005} prediction of the average comoving radius enclosing $90\%$ of the stellar mass in protocluster evolution with redshift, which are 13.2 ($\sim 9500$), 14.4 ($\sim 12500$) and 14.4 ($\sim 12500$) cMpc (cMpc$^3$) for the Slug, J0819 and Fabulous protoclusters, respectively. We will refer to this volume as the “predicted” volume. For the estimate of the AGN space densities of all protoclusters, we removed the central QSOs, since they were used to select the structures in the first place, and thus their inclusion would bias the XLF estimation.
\par
\begin{figure*}[]
\centering
{\includegraphics[width=0.5\textwidth]{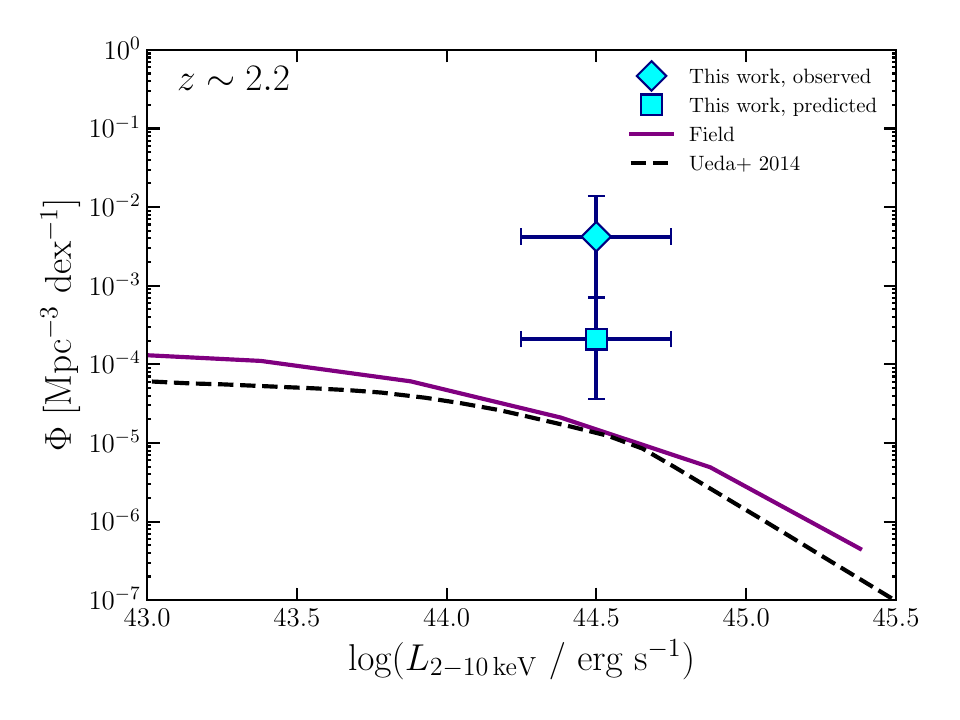}}
\hspace{-0.2cm}
{\includegraphics[width=0.5\textwidth]{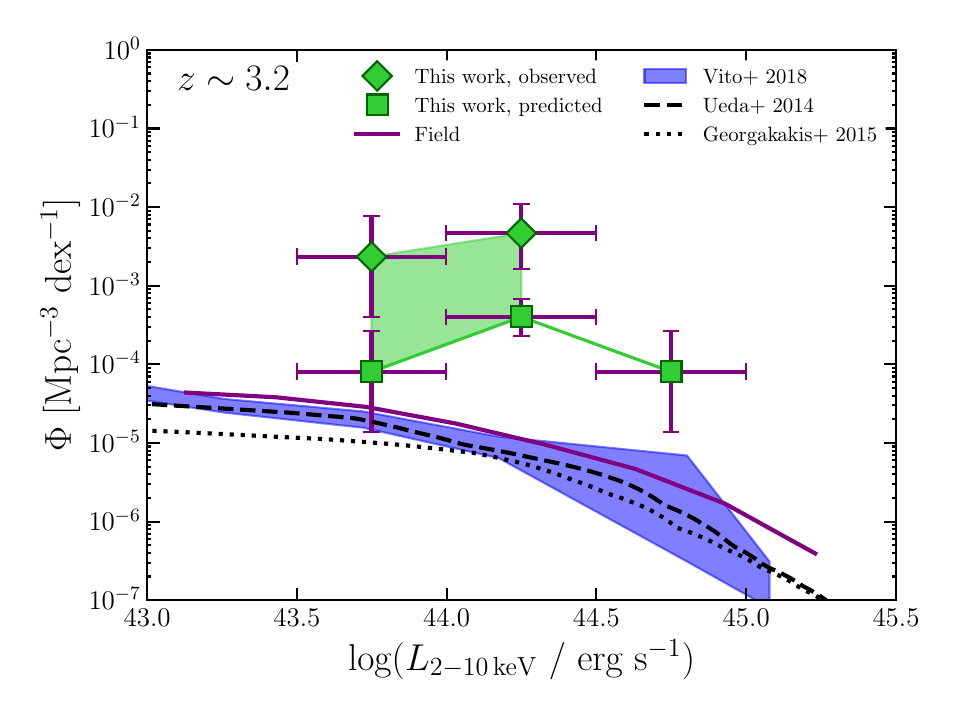}}
\caption{X-ray luminosity functions in two different redshift bins ($z \sim 2.2$ and $z \sim 3.2$) and in luminosity bins of 0.5 dex. Our data are represented by the cyan and green points (with different markers representing different assumptions on the co-moving volume). As a comparison, we report the XLFs for the field environment at similar redshifts. The purple solid line is the XLF for field AGN by \citet{gilli2007}. The dashed lines are the XLF by \citet{ueda2014} at $z\sim2.2$ and $z\sim3.2$. The dotted line is the XLF by \citet{georgakakis2015} and the purple shaded area is the XLF estimated by \citet{vito2018}}.
\label{fig:XLF}
\end{figure*}

 We estimated the XLF in the two redshift bins by assuming the volumes defined before. In this way, we can compare our results with the XLFs in blank fields from literature \citep[][]{lafranca2005,gilli2007,ueda2014,georgakakis2015,vito2018}. The results are shown in Figure \ref{fig:XLF}. We find that the XLF for protoclusters is higher than what is observed in fields by a factor $\times\, 10$ if we consider the protocluster radius predicted by simulations, while considering the “observed” volume this difference is up to two order of magnitudes.
\par
\begin{figure*}[]
\centering
{\includegraphics[width=0.5\textwidth]{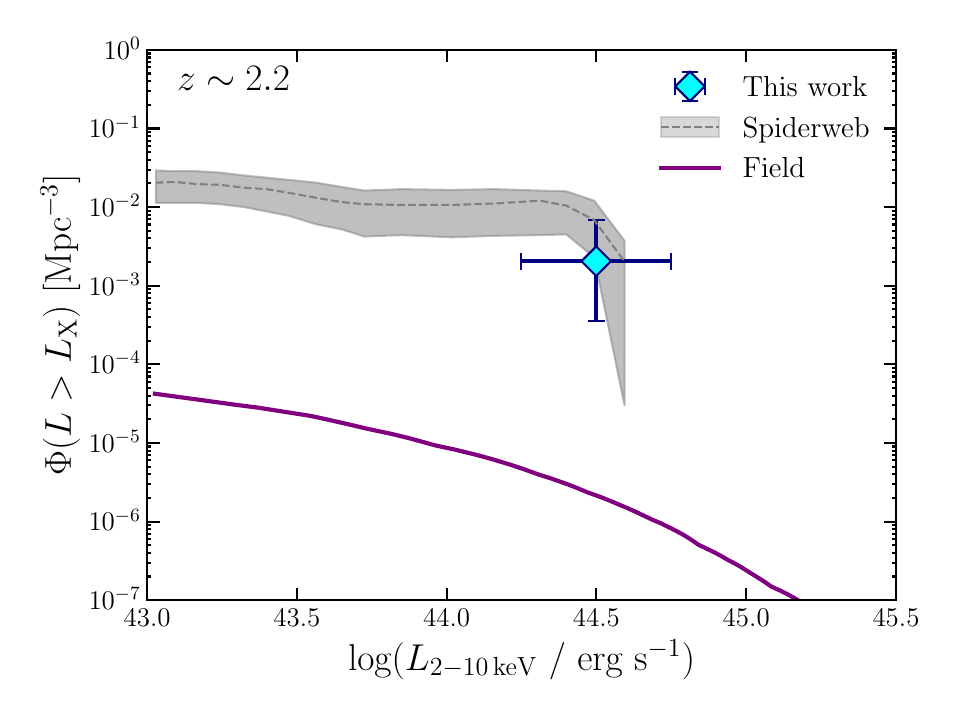}}
\hspace{-0.2cm}
{\includegraphics[width=0.5\textwidth]{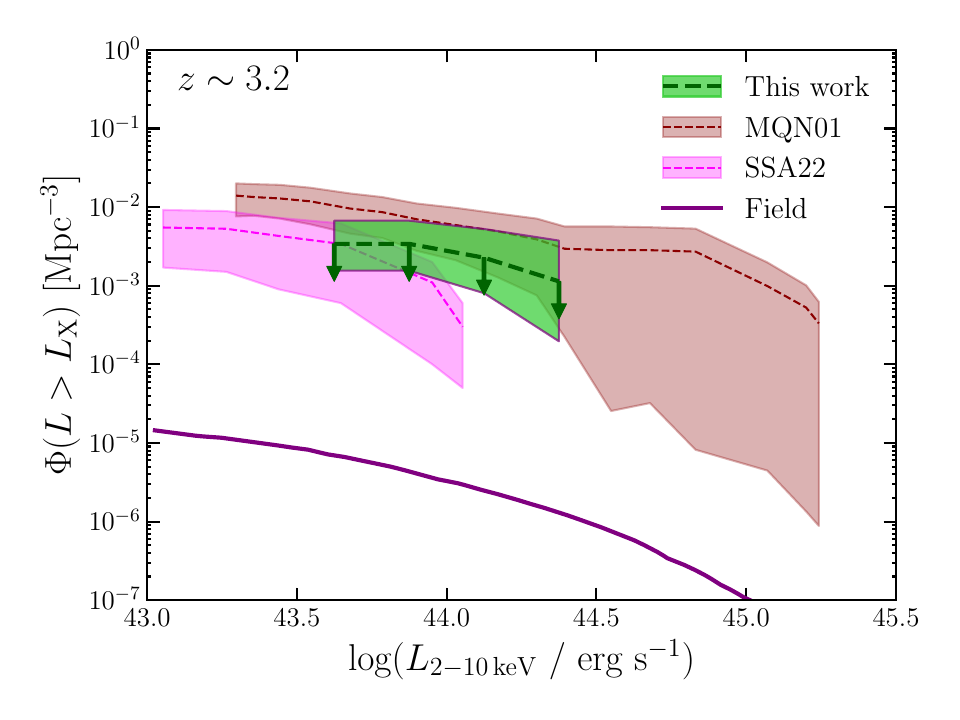}}
\caption{Cumulative space density in two different redshift bins ($z \sim 2.2$ and $z \sim 3.2$). Our data are represented by the cyan point and green area (where the uncertainties are computed according to \cite{gehrels1986errors}). As a comparison, we report the cumulative space densities of other protoclusters at similar redshifts. The gray shaded region (left panel) is the cumulative space density for the Spiderweb protocluster \citep[see][]{tozzi2022a,travascio2025}, while the red and pink shaded areas (right panel) are the MQN01 and SSA22 by \citet{travascio2025}. The purple solid line is the cumulative space density for field AGN by \citet{gilli2007}}. 
\label{fig:cumulative_XLF}
\end{figure*}

 Following the approach by \citet{travascio2025}, we also estimated the cumulative space density of the three protoclusters. In our case, we computed the cumulative AGN distribution and divided it for the comoving volume. To consistently compare our results to those by \citet{travascio2025}, we assumed the same volume (as defined before).
The results are shown in Figure \ref{fig:cumulative_XLF}, where the cumulative AGN space densities are reported  with errors obtained by using the Gerhels estimates for low statistics ($N_{\rm obj} < 10$). Due to the similar redshift, we joined the Fabulous and J0819 results into a single estimate, adding both the number of AGN and the volumes of the two structures. In the lower redshift bin, where the cumulative space density is obtained from the Slug sources, the cumulative space density is slightly lower than the space density found in the Spiderweb protocluster at similar redshift \citep[][]{tozzi2022a,travascio2025}, although consistent with it within the uncertainties, and it shows an excess of $\sim 3$ dex compared to fields. 
Similarly, in the second redshift bin the cumulative space densities significantly differ from what is found for fields \citep[][]{gilli2007}, with a $\sim 2-3$ dex discrepancy found using the same volume as \citet{travascio2025}. Even though most of the members of these structures are spectroscopically unconfirmed, we find an overall agreement with the results of \cite{travascio2025} for the MQN01 and SSA22 protoclusters. However, it must be taken into account that our observations are shallower than those by \citet{tozzi2022a} (700 ks) and \citet{travascio2025} (630 ks), leading to a shallower lower limit in the probed luminosities of the detected sources.
\par
\begin{figure*}[]
{\includegraphics[width=1.\textwidth]{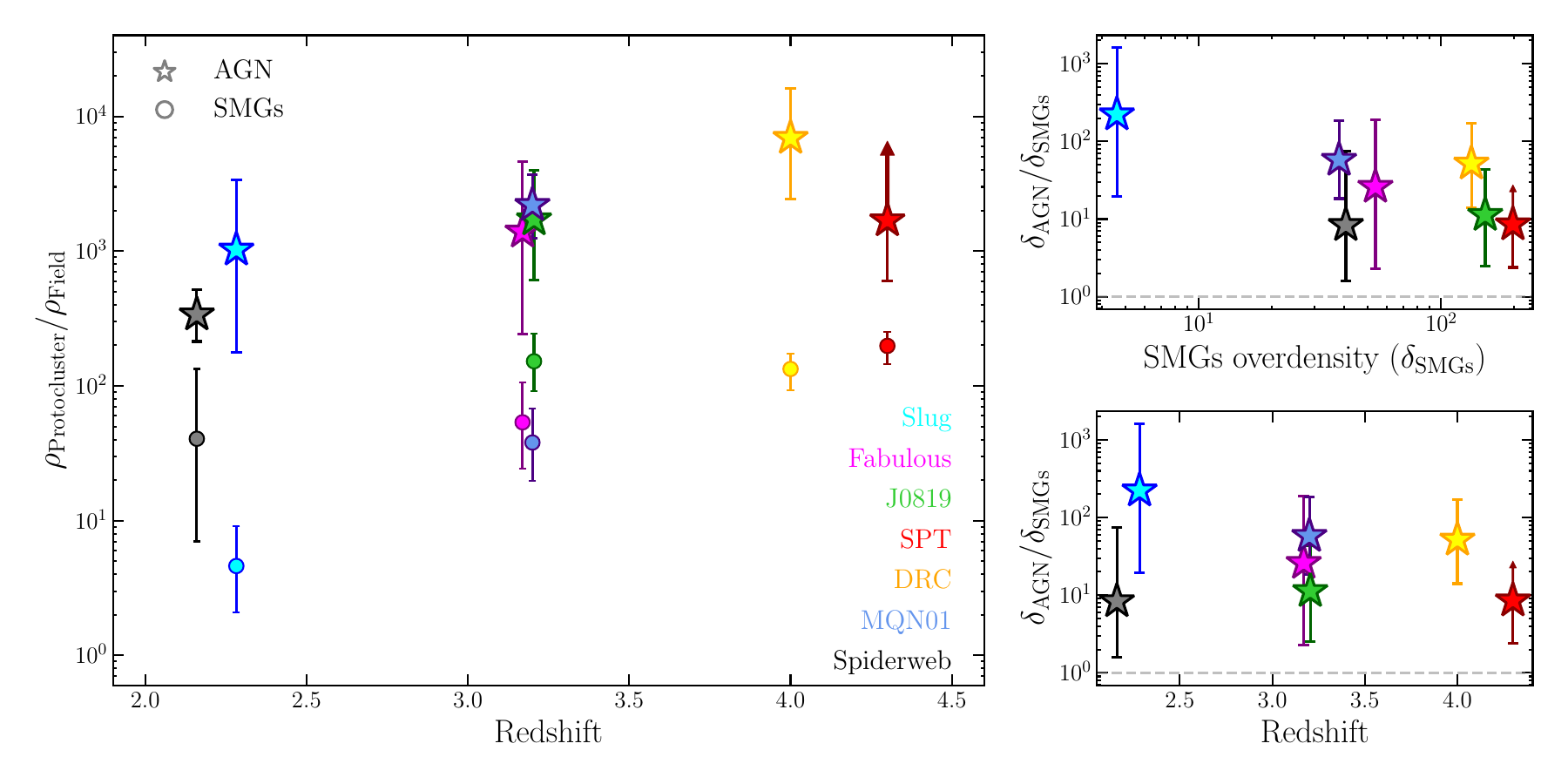}}
\caption{{\it Left panel:} Ratio between the number density of objects in the three protoclusters and in fields. We show this ratio for AGN (stars) and SMGs (circles). Errors are computed in the Poissonian assumption. The environment of AGN with respect to field in our targets is 1-2 dex larger than overdensity of the parent population of galaxies, impliying a direct environmental effect on the SMBH growth in these structures. We also show the ratios for other protoclusters at various redshifts \citep[][]{tozzi2022a,vito2020,vito2024,travascio2025}. {\it Right panel:} In the top-right panel we show the ratio between the AGN and SMGs overdensity as a function of the SMGs overdensity for the same protoclusters, while the bottom-right panel is the same but as a function of the redshift}.
\label{fig:AGN_density}
\end{figure*}

 To understand more deeply if the larger number of AGN observed in protoclusters reflects the higher density of galaxies or it is the product of faster SMBH growth, we compared the ratio between the number density of AGN in the protoclusters and in the field (i.e., AGN overdensity) with the same ratio computed for galaxies (i.e., galaxy overdensity). This analysis was conducted by assuming the “observed” volumes of the protoclusters. A similar protocluster-to-field ratio for AGN and galaxies would mean that the larger AGN number in protoclusters derive directly from the overdensity of galaxies. If instead the ratio for AGN is higher than the ratio for galaxies, we can conclude that the SMBH growth is directly enhanced in overdense regions of the Universe. To compute the aforementioned ratios, we assumed a XLF for AGN and IR-LF for SMGs, in fields. In particular, we assumed the \cite{gilli2007} XLF\footnote{\href{http://www.bo.astro.it/~gilli/index.html}{http://www.bo.astro.it/~gilli/index.html}} for AGN and the IR-LF by \cite{Traina2024sfrd} for SMGs. By integrating these curves we obtain the number density of that population of galaxies, which, multiplied by the volume of the protoclusters, give the expected number of field galaxies in that volume. The LFs have been integrated down to the luminosity of the faintest member of the corresponding population in our targets (thus taking into account the sensitivity of the observations).
For the SMGs, to convert from observed sub-mm fluxes into infrared $8-1000\,\mu {\rm m}$ luminosities, we assumed the SED of our SMGs to be similar to the median SED of the Automated mining of the ALMA Archive in COSMOS \citep[A$^3$COSMOS\footnote{\href{https://sites.google.com/view/a3cosmos}{https://sites.google.com/view/a3cosmos}};][]{liu2019a31,adscheid2024a3cosmos,Traina2024sfrd}. We then compare the expected number of sources with the observed ones in the protoclusters, estimating the AGN and SMGs protocluster-to-field ratios. We show the results of this comparison as a function of redshift in Figure \ref{fig:AGN_density} (left panel). For the Slug protocluster, the overdensity of AGN is higher (a factor $\gtrsim 100$) than the overdensity of SMGs. The discrepancy is smaller for the Fabulous and J0819 protoclusters, where the difference is of a factor $\sim 10$.
For comparison, we computed the same ratios for other protoclusters from the literature, allowing us to cover a broader redshift range. In particular, we considered the Spiderweb ($z \sim 2.16$), MQN01 ($z \sim 3.2$), DRC ($z \sim 4$) and SPT2349-56 ($z \sim 4.3$) protoclusters. The ratios were computed as for those in our sample. We find also for these protoclusters a significant discrepancy between the AGN and SMGs enhancements, suggesting the ubiquity of this phenomenon at all redshifts. In addition, an increasing trend with redshift seems to emerge, from these results, for the enhanced SMBH activity in protoclusters (note that for SPT we derived a lower limit due to the lower limit on the X-ray luminosity of the faintest member). However, the SMGs overdensity also is increasing with the redshift. Thus, to investigate whether a redshift evolution is present in the AGN-to-SMGs enhancement, we derived the ratios of the two overdensities for all the protoclusters (see Figure \ref{fig:AGN_density}, right panel). We find that all these structures have a AGN enhancement that is significantly higher than that of SMGs. In addition, this enhancement does not seem to change with the overdensity of SMGs (i.e., the larger density of AGN does not depend on the size of the overdensity).
In conclusion, our estimates of the volume densities, along with additional observational evidence from other works at similar or higher redshifts \citep[see e.g.,][]{tozzi2022a, travascio2025, vito2024} support a scenario in which the growth of SMBHs, and, as a consequence, the phenomena of active nuclei, is significantly favoured in overdense regions of the Universe, where more gas is available to fuel and infall into the nuclear region of galaxies. The discrepancy with field, moreover, is enhanced at higher X-ray luminosities, where protocluster XLFs show a roughly flat trend (Figures \ref{fig:XLF} and \ref{fig:cumulative_XLF}), without a significant decrease of the volume density, once again supporting the scenario of the faster formation of SMBH of larger masses. 

\subsection{Contribution of cooling radiation on ELAN emission}\label{elane_powering}

\begin{figure}[]
{\includegraphics[width=.5\textwidth]{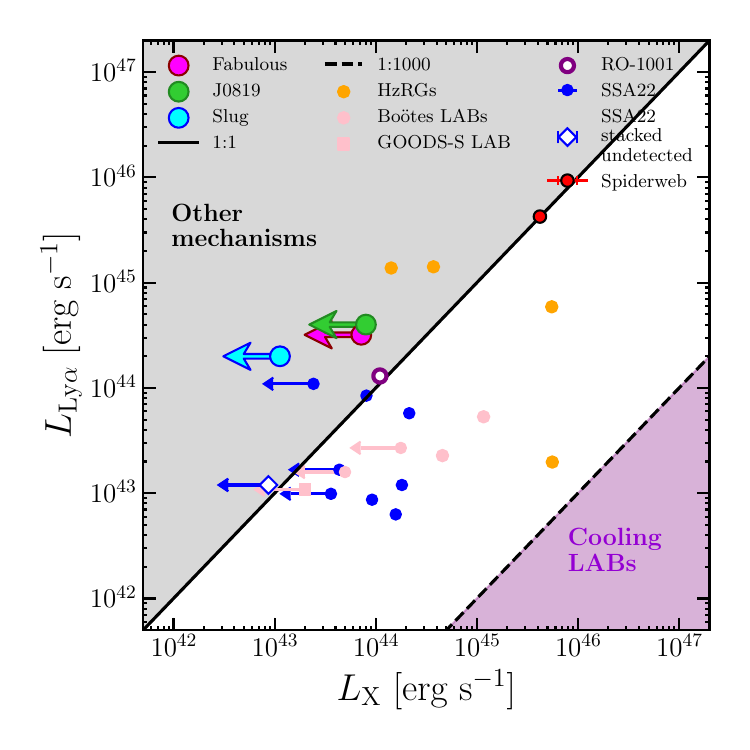}}
\caption{X-ray vs Ly$\alpha$ luminosity of the extended emission from the ELAN: magenta, green, and cyan circles represent the Fabulous, J0819, and Slug nebular, respectively. We compare them with values for other extended Ly$\alpha$ emission in the literature. Blue and white points are SSA22 extended Ly$\alpha$ emission reported by \citet{geach2009}; orange circles are extended Ly$\alpha$ emission surrounding HzRGs from \citet{reuland2003}; extended Ly$\alpha$ emission in the Boötes and GOODS-S fields are shown as pink circles and squares, respectively \citep[][]{yang2009,nilsson2006}; the red circle is the Spiderweb extended Ly$\alpha$ nebula by \citet{tozzi2022b} and the withe circle is the LAB by \citet{daddi2021}. The black solid line represents the 1:1 relation \citep[][]{geach2009}, while the black dashed line is the $L_{\rm X} = 1000 L_{\rm Ly\alpha}$ relation \citep[][]{cowie1980,bower2004}. The purple region (lower left corner) corresponds to the scenario in which the main powering channel of the extended Ly$\alpha$ emission is the gravitational cooling. The gray area (upper right corner), instead, represents the region in which the cooling cannot be the fueling mechanism of the nebula.}
\label{fig:LX_Llya}
\end{figure}

Since we do not detect any significant extended X-ray (1-2 keV) emission co-spatial with the ELAN in the Slug, J0819 and Fabulous protoclusters, in this section, we extract a $3\sigma$ upper limit on the observed flux and intrinsic X-ray luminosity for the diffuse emission. In the hypothesis that some of the Ly$\alpha$ emission is not powered by the bright QSOs, e.g., as a consequence of photo-ionization or Ly$\alpha$ photon-pumping \citep[see e.g.,][for a review]{cantalupo2017}, even though the ionizing power of the AGN in the protoclusters is more than 1 dex higher than that required to power the nebula \citep{umehata2019}, these estimates can provide constraints on additional or alternative powering mechanisms such as the so called Ly$\alpha$ cooling \citep[see e.g.,][]{Rosdahl2012} which has been advocated in the past to explain extended Ly$\alpha$ emission without clear photoionization sources, or Ly$\alpha$ Blobs \citep[LAB; see e.g.,][]{geach2009,daddi2021}. In particular, in the assumption that a significant fraction of the gas is not photoionized by the quasar (e.g. because the quasar is young or the opening angle of the emission is very small), we can estimate the expected $L_{\rm X} / L_{\rm Ly \alpha}$ ratio produced by a cooling flow in collisional ionization equilibrium. Following the works of \citet{bower2004, geach2009} the conservation of angular momentum in subsonic cooling flows would imply $L_{\rm X} / L_{\rm Ly \alpha} > 1000$ if Ly$\alpha$ emission is produced by Ly$\alpha$ collisional excitation of energetic electrons powered by gravitational heating, the so called “Ly$\alpha$ cooling”. On the other hand, if $L_{\rm X} / L_{\rm Ly \alpha} < 1$, we can be confident that in the above assumptions that the majority of the emission of the nebula is unrelated to Ly$\alpha$ cooling. We derived the $3\sigma$ upper limits on the ELAN X-ray fluxes and luminosities in regions surrounding the central QSOs. We extracted the X-ray spectra and ancillary files from a annular region (inner radius $=2^{\prime\prime}$, outer radius $=4^{\prime\prime}$) centered on the QSO. In this case, since the emission is not point-like, we runned \texttt{specextract} selecting the option \texttt{weight=True}, allowing for the weighting of the ARF and RMF associated with the extracted spectra. We then obtained the $3\sigma$ upper limits on the $1-2$ keV fluxes and corresponding rest-frame $\sim 4-8$ keV intrinsic luminosities to compare directly with the results of \citet{geach2009}, by scaling the upper limits on the net counts in these regions assuming a model with $\Gamma \sim 1.8$ as intrinsic spectrum. We constrain the X-ray luminosities of the ELANe to be $< 7.2 \times 10^{43}$ erg s$^{-1}$, $< 8 \times 10^{43}$ erg s$^{-1}$ and  $< 1.1 \times 10^{43}$ erg s$^{-1}$ for the Fabulous, J0819 and Slug, respectively. Considering their Ly$\alpha$ luminosities of $L_{\rm Ly\alpha} \sim 3.2 \times 10^{44}$ erg s$^{-1}$, $L_{\rm Ly\alpha} \sim 4 \times 10^{44}$ erg s$^{-1}$ and $L_{\rm Ly\alpha} \sim 2 \times 10^{44}$ erg s$^{-1}$, we estimate $L_{\rm X} / L_{\rm Ly\alpha} < 0.2, 0.2 \,{\rm and}\, 0.06$, respectively. In Figure \ref{fig:LX_Llya} we compare our results with those from the literature. Three different regions are highlighted: the gray one, with $L_{\rm X} < L_{\rm Ly\alpha}$, is populated by sources for which a cooling explanation for the ELAN emission is unlikely; the purple region, where $L_{\rm X} > 1000 \times L_{\rm Ly\alpha}$, in which the ELAN emission is consistent with gravitational cooling processes; finally, the white region is populated with sources with intermediate values of $L_{\rm X} / L_{\rm Ly\alpha}$. Out of 29 extended Ly$\alpha$ emission regions in the SSA22 protocluster, only 5 are X-ray detected, while for the remaining 24, an upper limit is derived by \cite{geach2009}. In particular, the 5 X-ray detected extended Ly$\alpha$ emission belong to the intermediate region, thus no strong constraint can be placed on the origin of their emission. The stacked X-ray emission of the SSA22 extended Ly$\alpha$ emission is instead on the gray region of the plot, thus their main powering mechanism cannot be gravitational cooling. Other extended Ly$\alpha$ emission, in the Boötes and GOODS-S fields are either in the intermediate region or undetected in X-rays \citep[][]{nilsson2006,yang2009}. Similarly, four high redshift radio galaxies (HzRGs) studied by \citet{reuland2003} are detected in the X-rays, but still in the gray or white areas. We show the Ly$\alpha$-Xray luminosities comparison also for the Spiderweb protocluster \citep[][]{tozzi2022b}, where an extended diffuse emission is well detected in the X-rays. In this case, we consider the total (i.e., including contaminant sources like the jets) for both the X-ray and Ly$\alpha$ luminosity, as a value of $L_{\rm Ly\alpha}$ is not available. Its X-ray luminosity is almost $5 \times 10^{45}$ erg s$^{-1}$, compatible with the luminosity of the total Ly$\alpha$ emission, placing the Spiderweb in the middle region where cooling cannot be ruled-out.
As most of the undetected sources, the three ELANe in our analysis lie on the gray region, where, as expected, gravitational cooling is not the mechanism responsible for the nebula emission.

\section{Conclusions}\label{sec:conclusion}
In this paper we presented deep \cha observations ($\sim 190$ ks, $\sim 240$ ks, $\sim 280$ ks) of three protoclusters (Slug, Fabulous and J0819) at $z \sim 2$ and $z\sim3$. The three protoclusters show evidences of large gas reservoirs in their inner regions, with enormous Ly$\alpha$ nebulae extending for several kiloparsecs. Here we summarize the results from the dedicated X-ray analysis on the point-like sources as well as for the (lack of) diffuse emission surrounding the central QSO. We also discuss the implications of our results on the fast growth of SMBHs in overdense regions of the Universe.

\begin{itemize}
    \item We detected a total of 11 X-ray AGN in the three protoclusters, 6 of which were not already known as AGN. Most of the detected AGN (9/11) are hosted in SMGs, while just 5 (which are the QSOs) are also LAEs.
    
    \item The main properties of the nuclear emission of these detected sources were obtained either via spectral analysis (for AGN with more than 30 net counts) or via their hardness ratios. The bright QSOs in the center of the ELANe presents similar photon indices and are unabsorbed. Their $2\,-\,10$ X-ray luminosities are larger than $10^{45}$ erg s$^{-1}$, making them very bright in the X-rays. In particular, QSO\_J0819 is the most luminous source in our sample, with $L_{\rm X} \sim 6 \times 10^{45}$ erg s$^{-1}$. The other AGN do not exhibit peculiar properties, except for one Fabulous AGN (Fab-33) that is Compton-thick candidate.

    \item We investigated the presence of potential diffuse emission, co-spatial with the ELANe, by comparing the observed radial profiles around the central QSOs with the simulated PSFs of each image. The emission around the QSOs is consistent with the shape of the Chandra PSF, ruling out a significant contribution from extended emission in our data. Nevertheless, we measured an upper limit on the X-ray luminosity around the QSOs, wich enabled us to rule out a cooling scenario for the supply of the Ly$\alpha$ nebulae.

    \item In good agreement with recent studies of the protoclusters XLF, we find the Slug XLF (at $z \sim 2.2$) and the Fabulous and J0819 XLF (at $z \sim 3.2$) to be $\sim 1-2$ orders of magnitude higher than the population of AGN in fields, confirming the larger number of AGN in these overdense environments, where we find the AGN fraction to be the $10\,-\,20\, \%$.

    \item This higher density of AGN cannot be ascribed just to the higher number of galaxies in the protocluster. Indeed, we find that the ratio of protocluster-to-field AGN is significantly higher than the same ratio computed for SMGs using homogeneous flux limits, supporting the scenario in which overdense, gas-rich environments enhance directly the growth of SMBHs in the member galaxies.    
\end{itemize}

In the next years, several currently operating and future facilities (e.g., Roman, Euclid, SKA) will be of fundamental importance in the detection and characterization of protoclusters at high redshift. Their contribution, combined with the potential of the AXIS probe will significantly improve our knowledge on the AGN content of such structures, allowing us to understand how these environments favor the formation of the most massive SMBHs in the Universe.

\begin{acknowledgements}
AT, FV, CV, PT, RG acknowledge support from the “INAF Ricerca Fondamentale 2023 – Large GO” grant. C.-C.C. acknowledges support from the National Science and Technology Council of Taiwan (111-2112-M-001-045-MY3), as well as Academia Sinica through the Career Development Award (AS-CDA-112-M02). AP acknowledges support from Fondazione Cariplo grant no. 2020-0902. SC gratefully acknowledges support from the European Research Council (ERC) under the European Union’s Horizon 2020 Research and Innovation programme grant agreement No 864361.
\end{acknowledgements}

\bibliographystyle{aa}
\bibliography{1biblio}

\begin{appendix}
\section{X-ray detections}\label{app:cutouts}

\begin{figure*}[]
\centering
{\includegraphics[width=1.\textwidth]{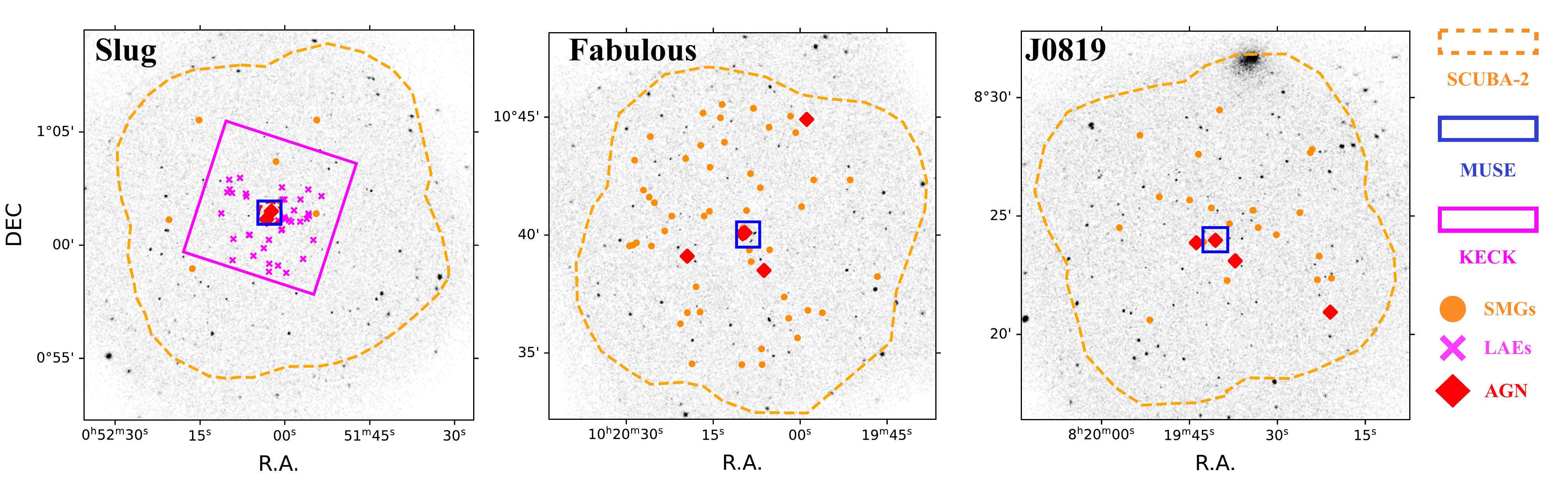}}
\caption{\cha X-ray maps (gray scale) of the three fields, centered on the aim point. The coverage of ancillary data is superimposed as colored contours (orange, blue, red and magenta for SCUBA-2, MUSE, ALMA and KECK I, respectively).}
\label{fig:maps}
\end{figure*}

\begin{figure*}[]
\centering
{\includegraphics[width=1.\textwidth]{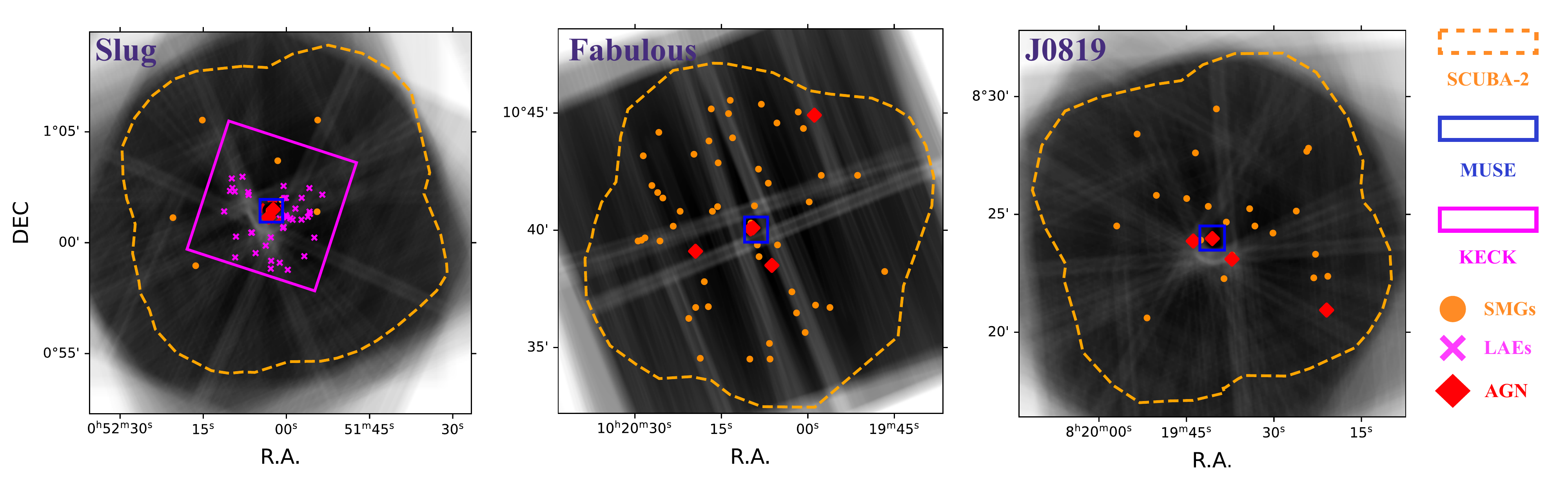}}
\caption{Same as Figure \ref{fig:maps} but for the exposure maps.}
\label{fig:expmaps}
\end{figure*}

\begin{figure*} 
\centering
 \includegraphics[width=0.7\textwidth]{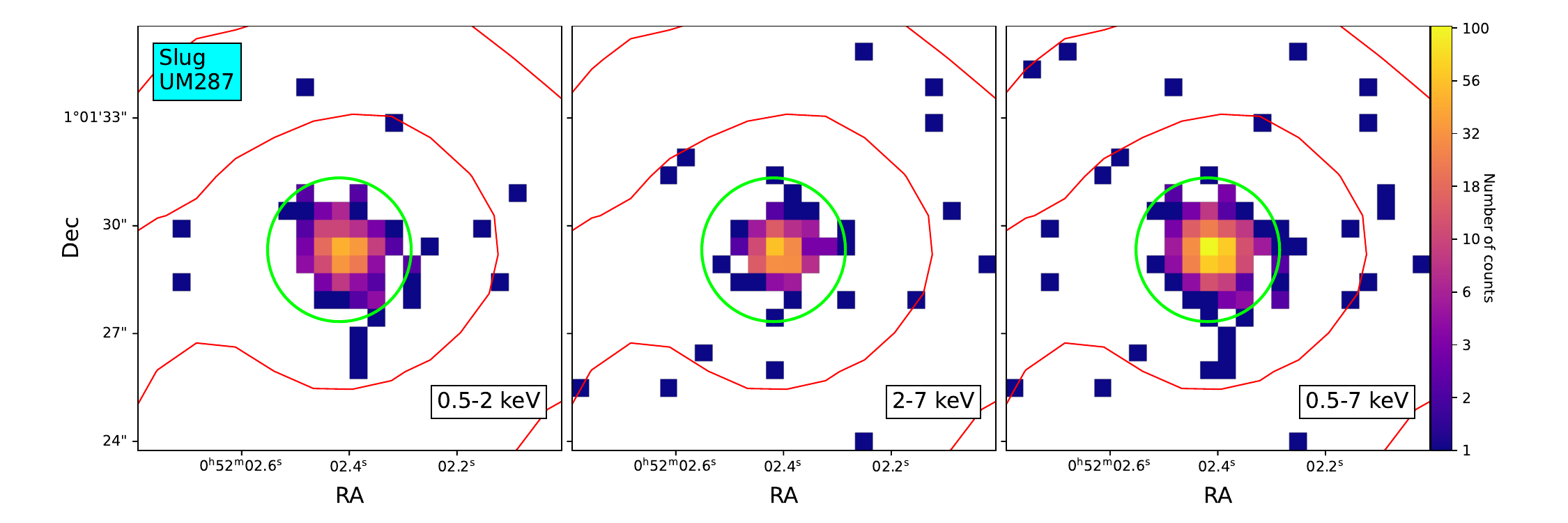}\\
 \includegraphics[width=0.7\textwidth]{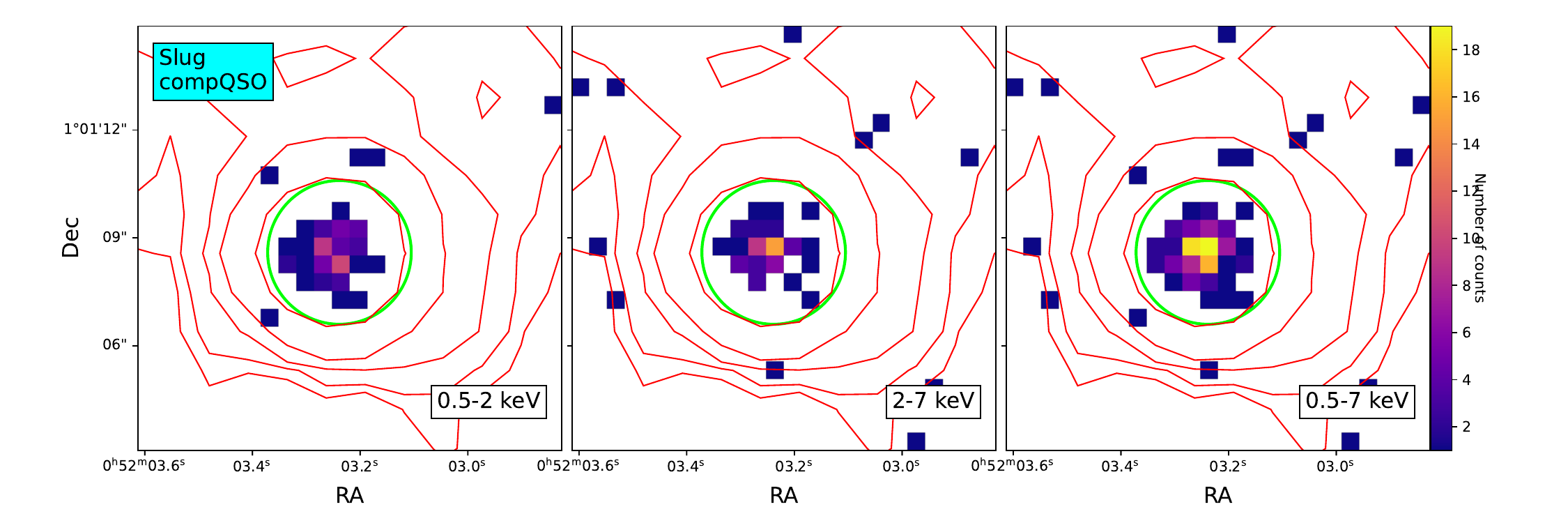}\\
 \includegraphics[width=0.7\textwidth]{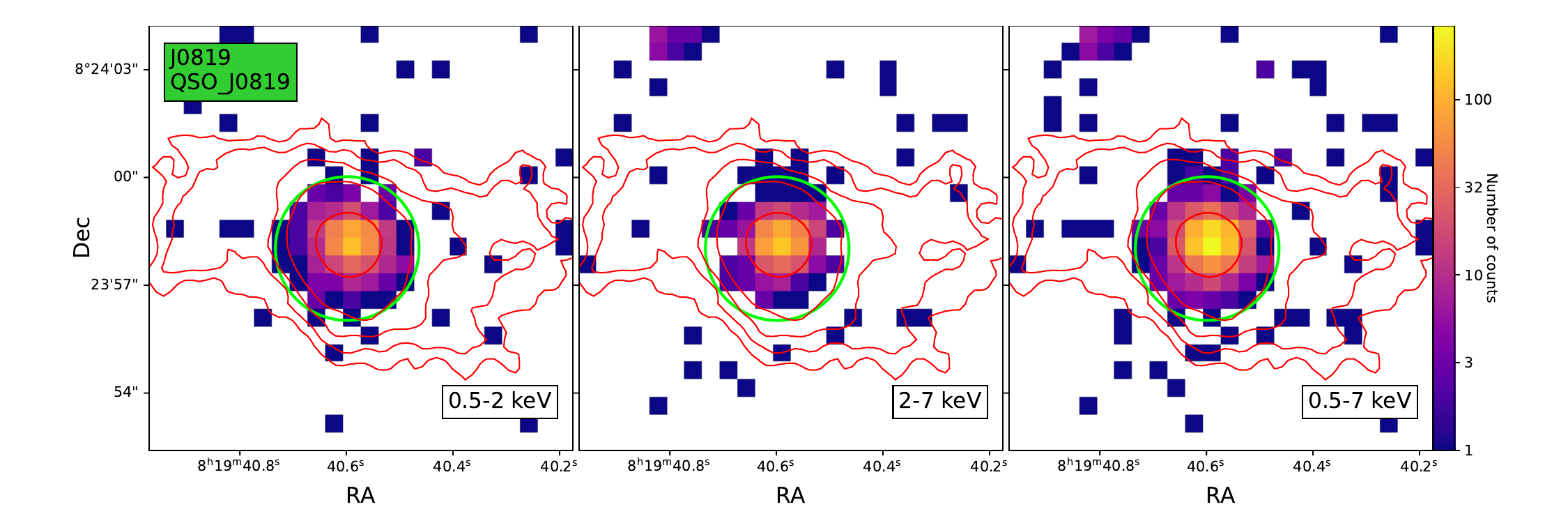}\\
 \includegraphics[width=0.7\textwidth]{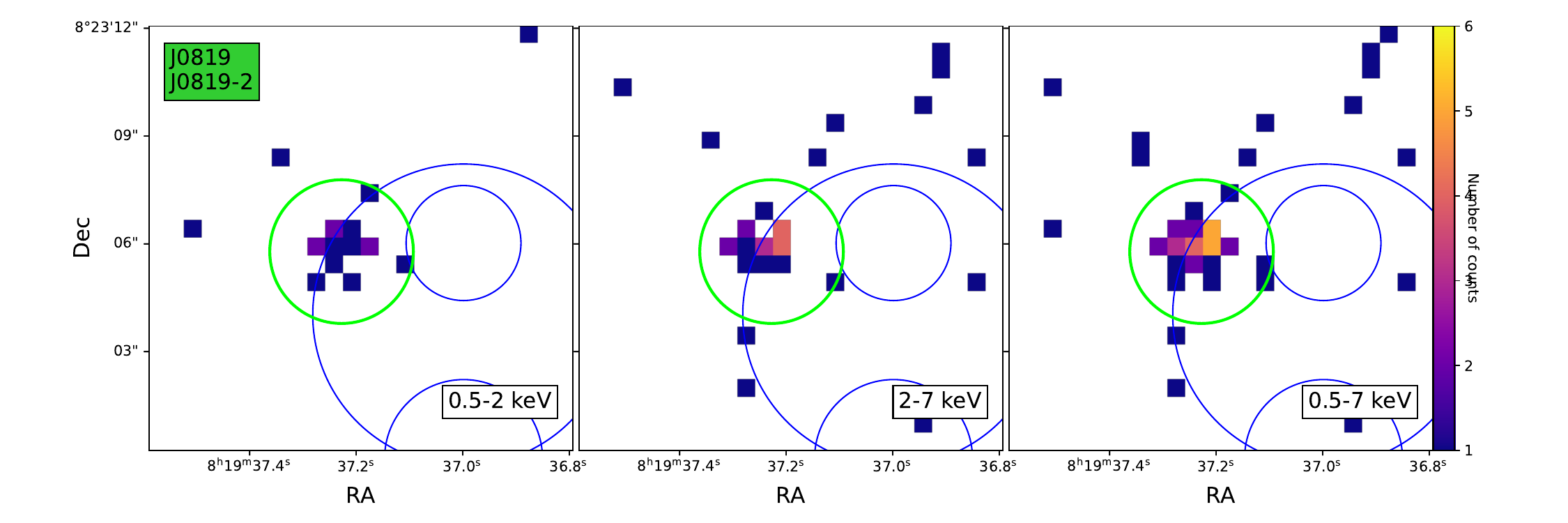}\\
 \includegraphics[width=0.7\textwidth]{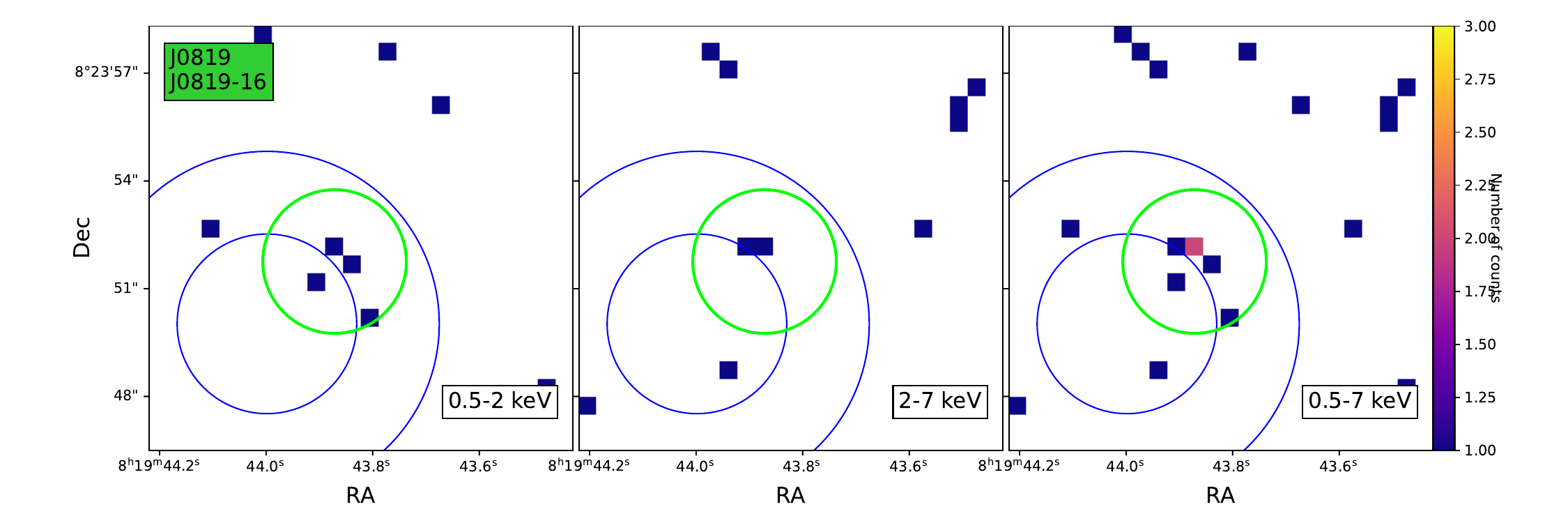}\\
 \includegraphics[width=0.7\textwidth]{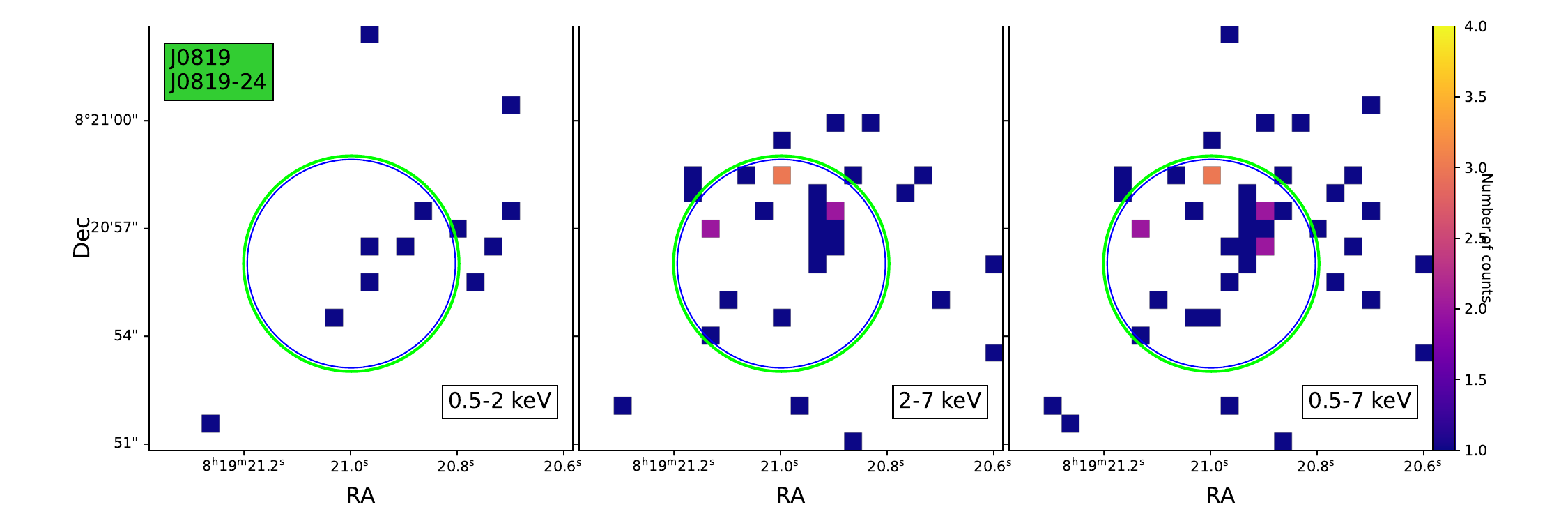}
 \caption{Cutouts of the X-ray detected AGN in the three protoclusters, in each band (soft, hard, full). Data are color coded by number of counts, while the green, blue and red contours are the extraction region, the prior position and the Ly$\alpha$ emission of the ELAN.}
  \label{fig:cutouts_all}
\end{figure*}

\begin{figure*} 
\centering
 \includegraphics[width=0.7\textwidth]{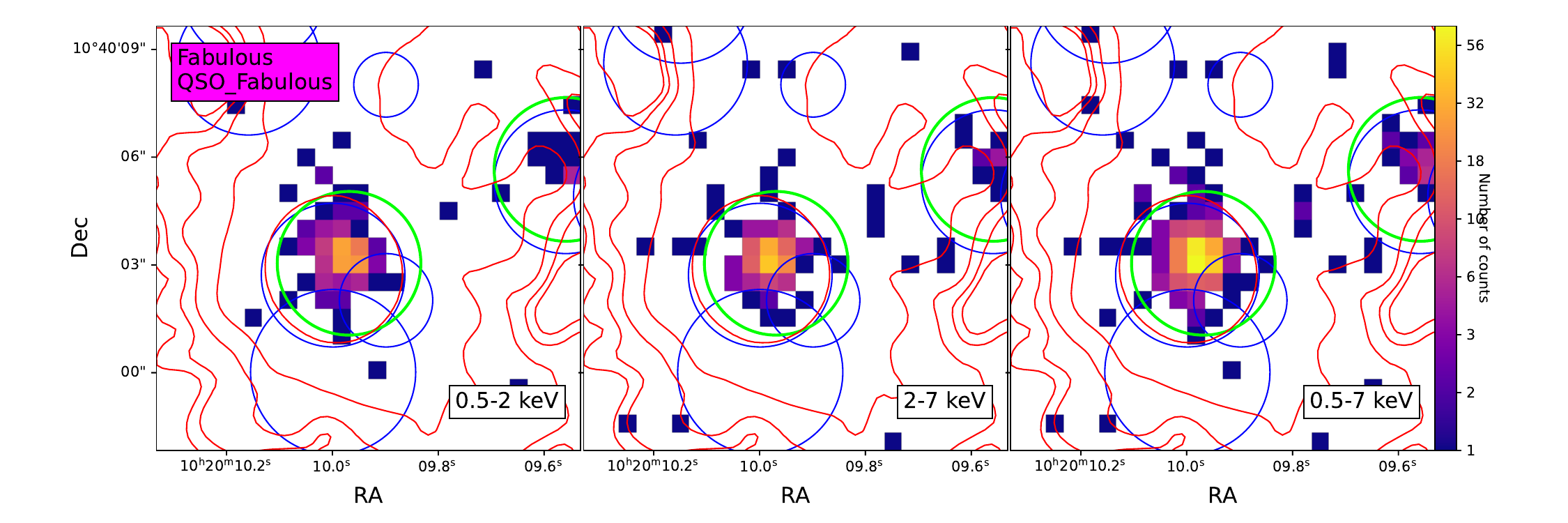}\\
 \includegraphics[width=0.7\textwidth]{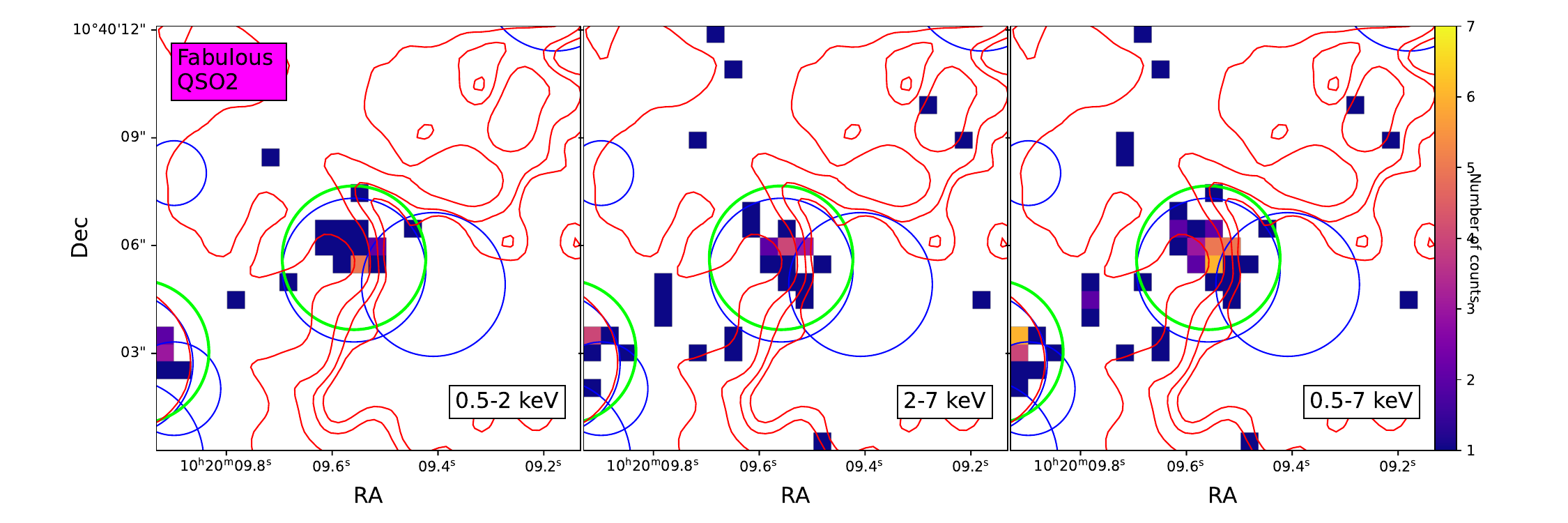}\\
 \includegraphics[width=0.7\textwidth]{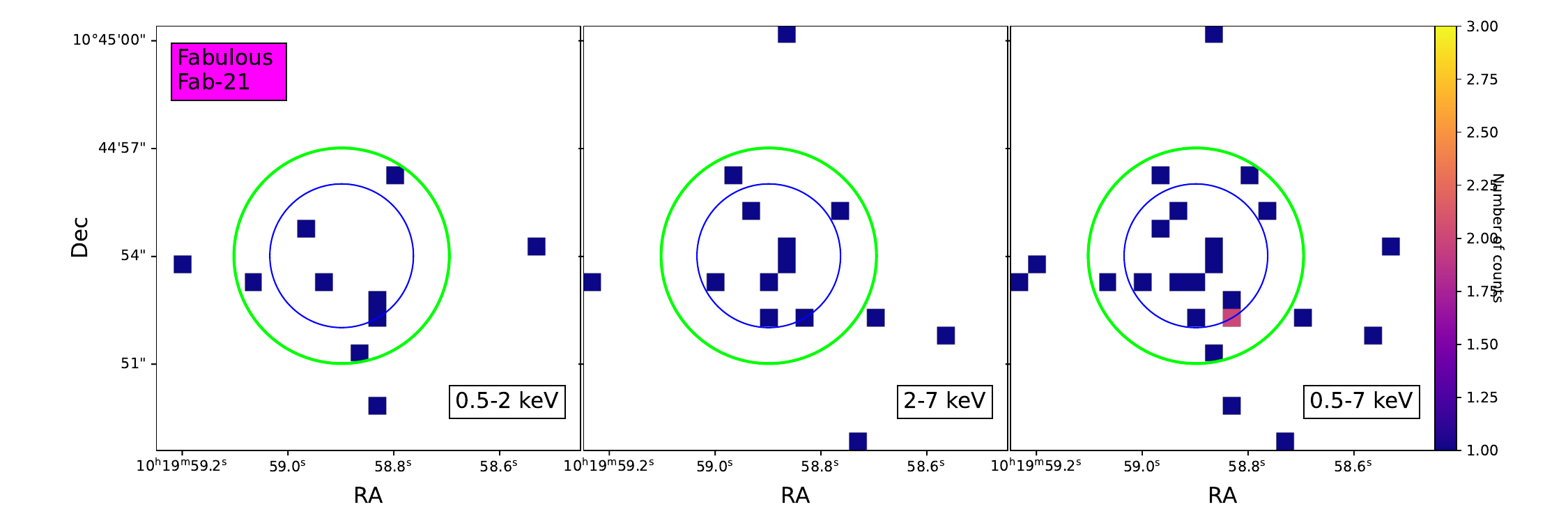}\\
 \includegraphics[width=0.7\textwidth]{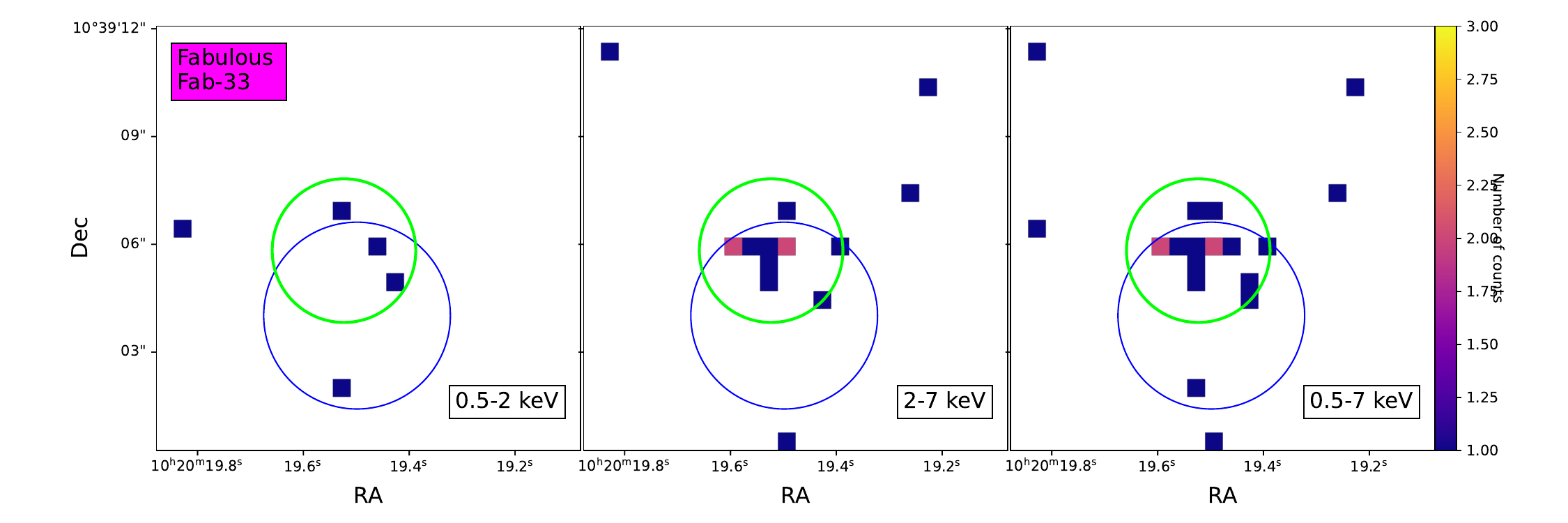} \\
 \includegraphics[width=0.7\textwidth]{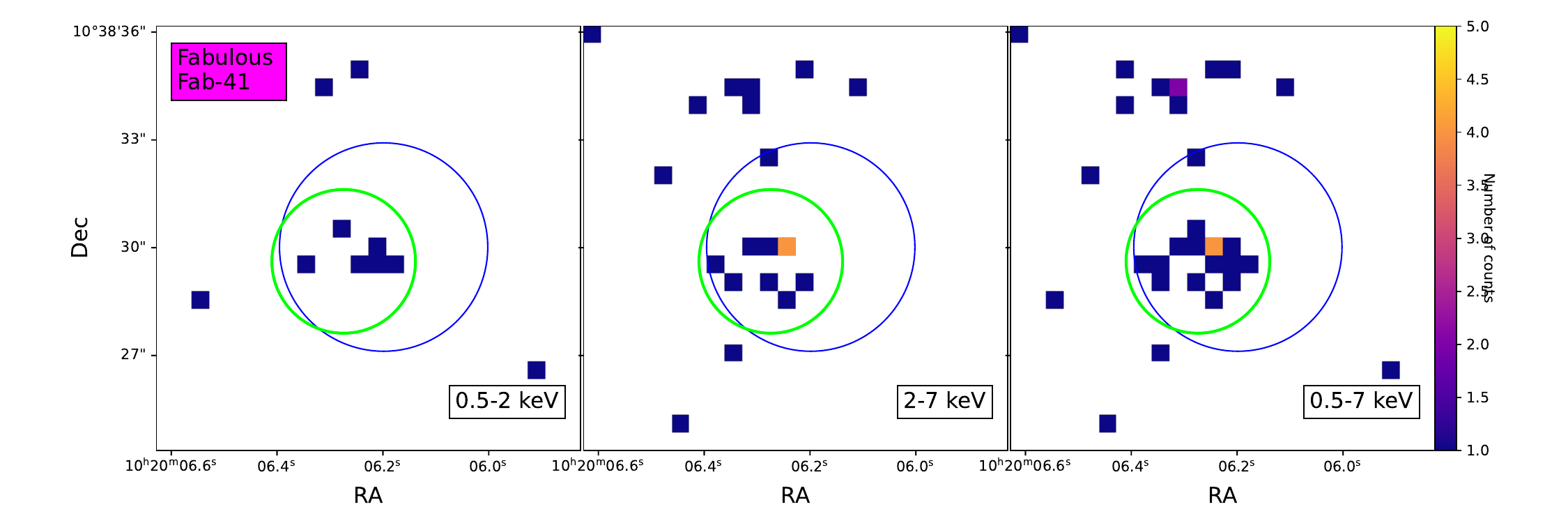} 
 \caption{Cutouts of the X-ray detected AGN in the three protoclusters, in each band (soft, hard, full). Data are color coded by number of counts, while the green, blue and red contours are the extraction region, the prior position and the Ly$\alpha$ emission of the ELAN.}
 \label{fig:cutouts_all2}
\end{figure*}

In this Appendix, we show the cutouts of X-ray detected AGN in the three protoclusters (the maps and exposure maps are shown in Figure \ref{fig:maps} and \ref{fig:expmaps}) and for the soft, hard and full bands (from left to right in Figures \ref{fig:cutouts_all} and \ref{fig:cutouts_all2}). We display as well the position of prior sources from the available catalogs (with a radius corresponding to the uncertainties on their position) and the extraction region centered on the X-ray source. In Table \ref{tab:general} we report the main informations on the protoclusters members.

\begin{table}[h!]
\renewcommand{\arraystretch}{1.5}
\caption{General informations on the three protoclusters.}
\begin{tabular}{ccccc}
\hline \hline
Protocluster & Redshift & $N_{\rm SMGs}$ & $N_{\rm LAEs}$ & $N_{\rm AGN}$ \\ \hline
Slug         & 2.2825   & 10 (3)             & 45 (2)            & 2 (2)                \\
Fabulous     & 3.1695   & 53 \phantom{00}                & 4  (4)            & 5 (2)               \\
J0819        & 3.205    & 25 (6)            & 1   (1)           & 4 (2)              \\ \hline
\end{tabular}
\begin{tablenotes}
   \small{ \item[*]{Name, redshift, number of SMGs, number of LAEs and number\\ of AGN members in the protoclusters. Values in brackets indicate\\ the number of secure members.}}  
\end{tablenotes}
\label{tab:general}
\end{table}

\section{X-ray spectra}\label{app:spectra}
\begin{figure*} 
\centering
 \includegraphics[width=0.3\textwidth]{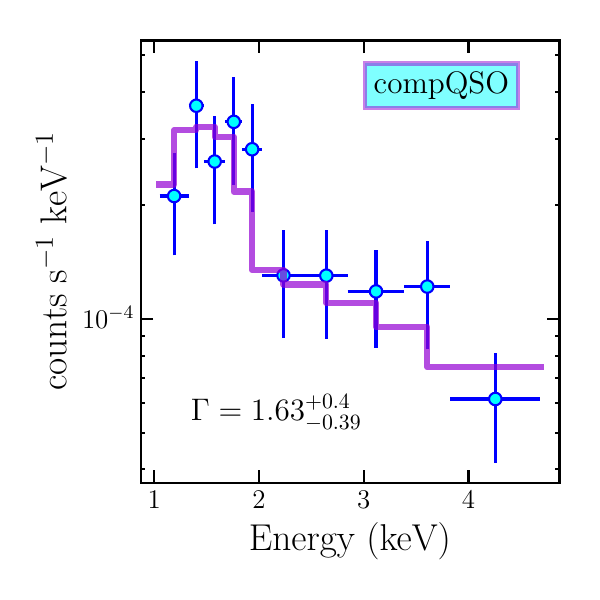}
 \includegraphics[width=0.3\textwidth]{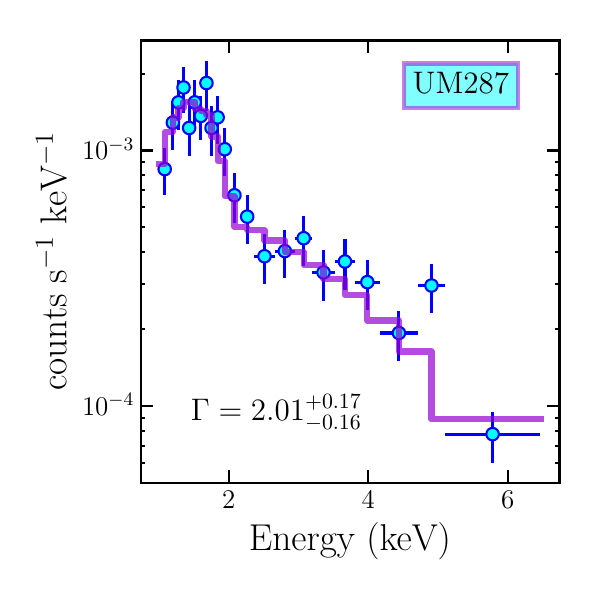}
 \includegraphics[width=0.3\textwidth]{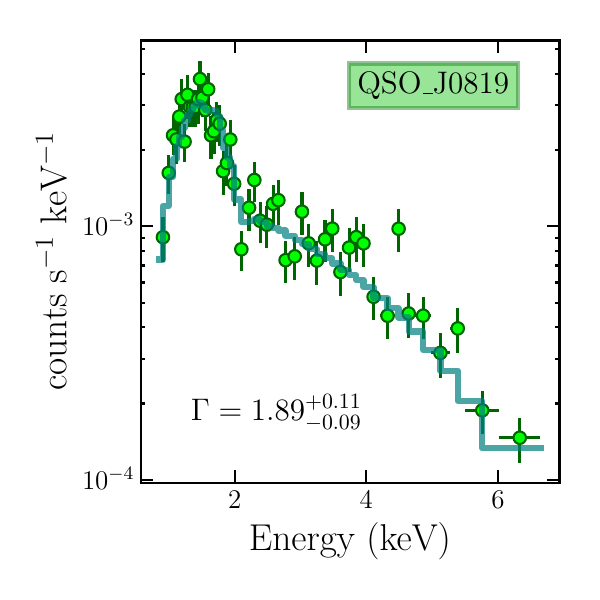}\\
 \includegraphics[width=0.3\textwidth]{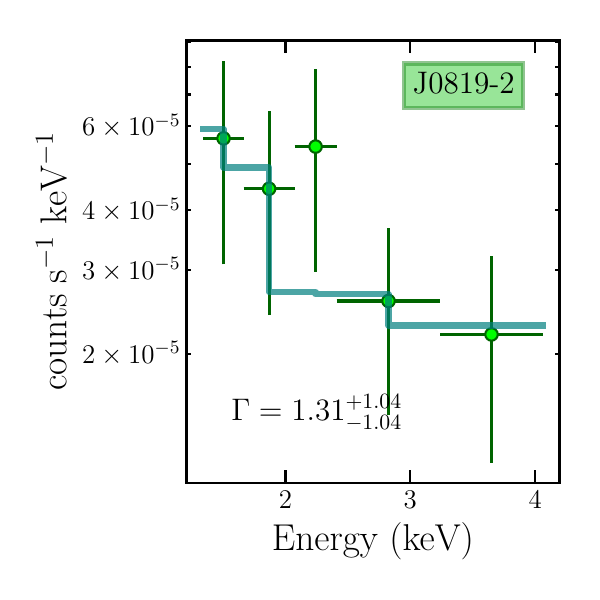} 
 \includegraphics[width=0.3\textwidth]{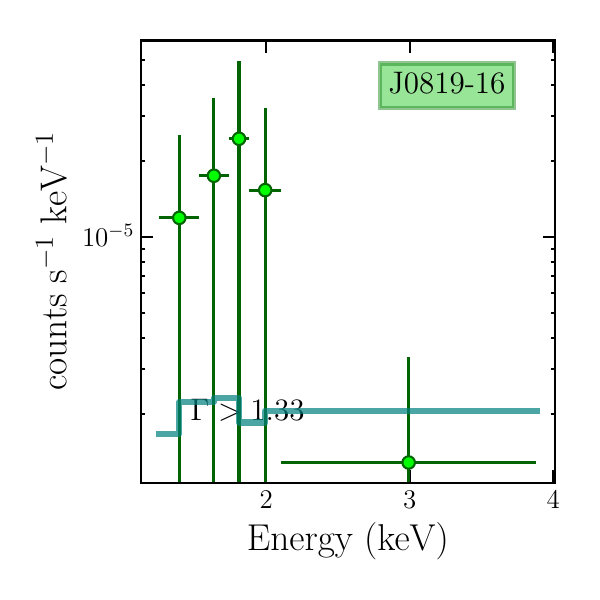} 
 \includegraphics[width=0.3\textwidth]{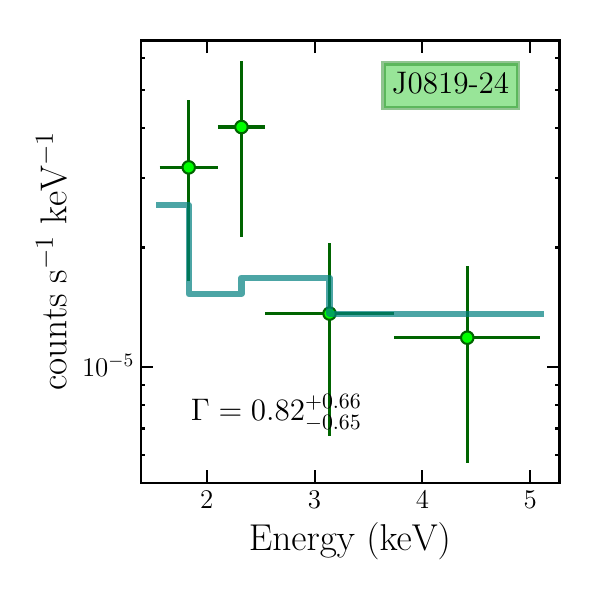} \\
 \includegraphics[width=0.3\textwidth]{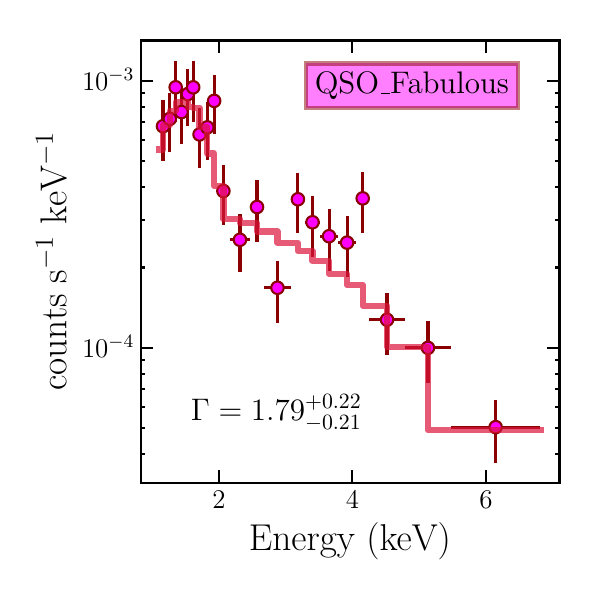} 
 \includegraphics[width=0.3\textwidth]{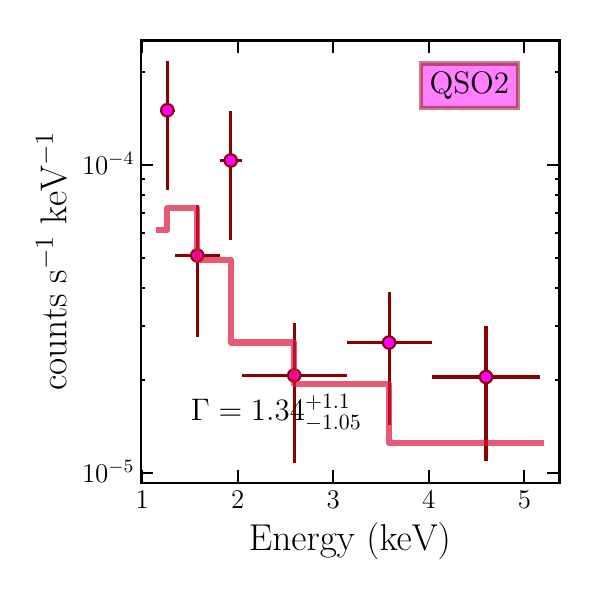} 
 \includegraphics[width=0.3\textwidth]{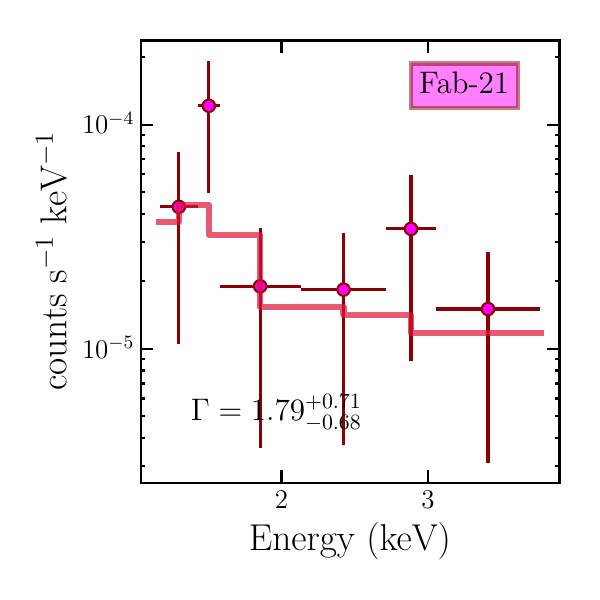} \\
 \includegraphics[width=0.3\textwidth]{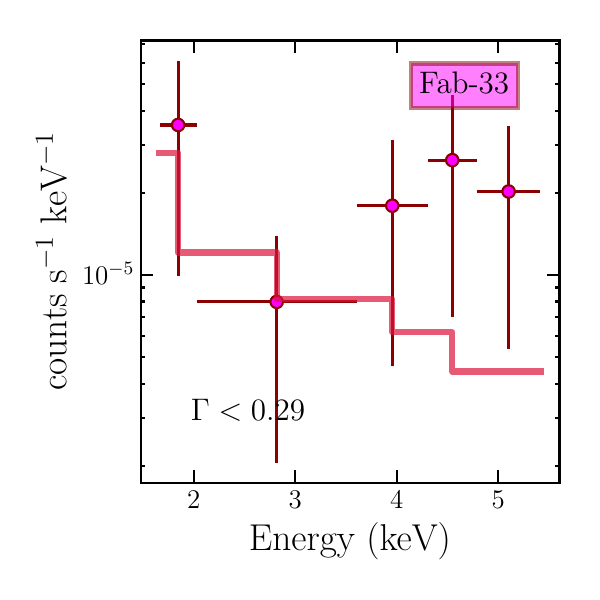} 
 \includegraphics[width=0.3\textwidth]{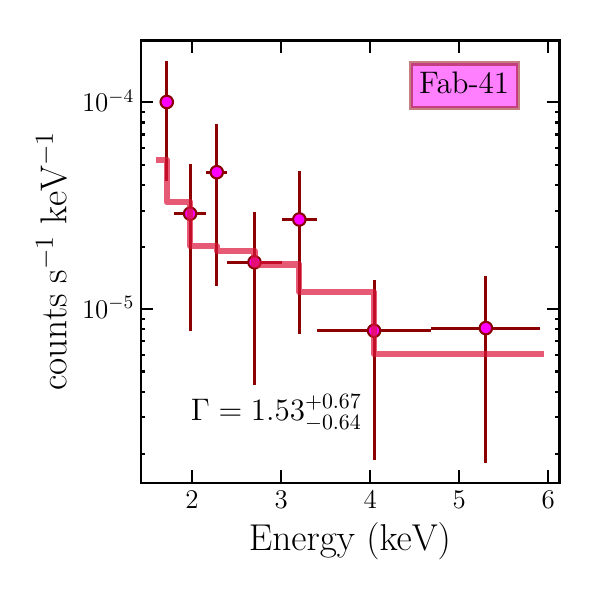} 
 \caption{X-ray spectra of the AGN detected in the three protocluster. The solid lines are the fit performed using an unabsorbed powerlaw. For sources with more than 30 net counts, the line represents the best fit, while for the other sources (J0819-16, J0819-24, Fab-21, Fab-33, Fab-41) is a power law with $\Gamma$ fixed to that found using the HRs.}
  \label{fig:spectra_all}
\end{figure*}

We performed the spectral analysis using \texttt{pyXspec} \citep[][]{gordon2021pyxspec} and report the results in Figure \ref{fig:spectra_all}. However, through the paper we use the results from this spectral analysis for sources with more than 30 counts in the full band.

\section{Slug optical photometry and SED fitting}\label{app:seds}

\subsection{Optical observations of the Slug protocluster}
We obtained new optical images of the Slug protocluster with the Large Binocular Camera (LBC) instrument mounted on the Large Binocular Telescope (LBT) with the \textit{U\_spec} (51 min), $B_{\mathrm{Bessel}}$ (20 min), $g_{\mathrm{SDSS}}$ (10 min), $V_{\mathrm{Bessel}}$ (15 min), $r_{\mathrm{SDSS}}$ (15 min), $i_{\mathrm{SDSS}}$ (15 min),  $z_{\mathrm{SDSS}}$ (37 min) filters. The reduction, astrometric and photometric calibration, and quality assessment of these data has been carried out at the LBC Survey Center\footnote{\url{http://lsc.oa-roma.inaf.it/}} in Rome, following standard LBC procedures and pipelines. We used the Graphical Astronomy and Image Analysis Tool (GAIA)\footnote{\url{https://astro.dur.ac.uk/~pdraper/gaia/gaia.html}} and the Source-Extractor \citep[][]{bertin1996sextractor} software to perform source detection and produce photometric catalogs.

\subsection{SED fitting properties of the Slug members}

\begin{figure*}[]
{\includegraphics[width=1.\textwidth]{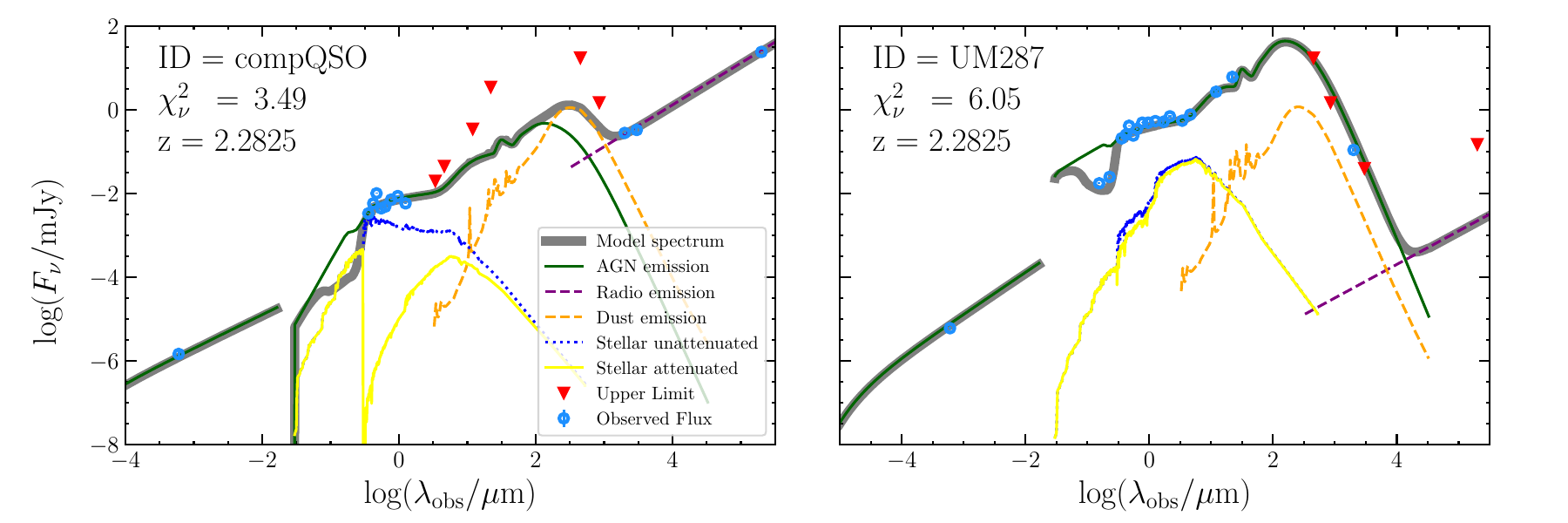}}
\caption{Best-fit SEDs of the central QSO (UM287) and companion QSO (CompQSO) in the Slug nebula. The best-fit model is shown in gray, while the individual contribution for each templates are shown in different colors: in green we show the AGN emission; the dust emission is displayed in orange; the stellar unattenuated and attenuated models are the dotted blue and solid yellow lines; the radio emission is highlighted in purple. Points and upper limit are shown as empty light-blue circles and red filled triangles, respectively.}
\label{fig:SED_fitting}
\end{figure*}

\begin{figure*}[]
{\includegraphics[width=.5\textwidth]{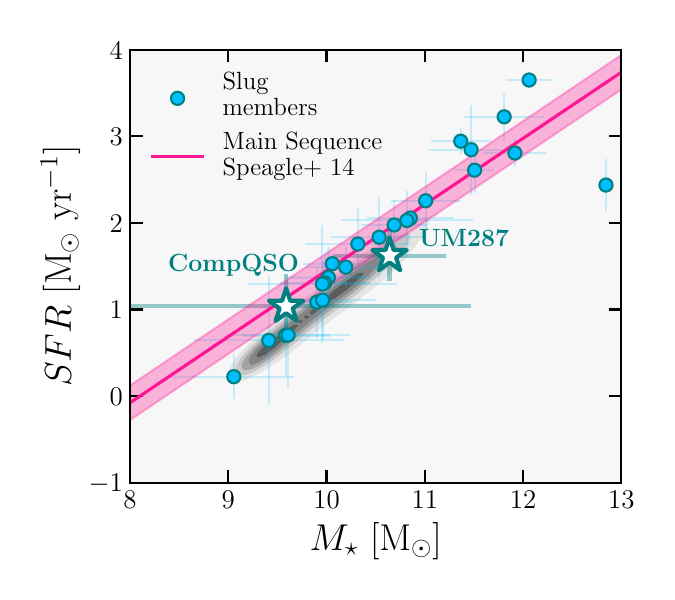}}
\caption{Main sequence of galaxies for the members of the Slug proto-\\cluster. The azure circles are the member galaxies. The white stars are\\ instead the two QSOs. The gray contours are the points from the \\XMM-SERVS \citep[][]{zou2022}. Finally, the magenta line represents the \\MS by \citet{speagle2014MS} a $z \sim 2.2$, plotted as a reference.}
\label{fig:MS_members}
\end{figure*}

In this section, we investigate the main physical properties (e.g., stellar mass, star formation rate) of candidate protocluster members (i.e., galaxies in the prior catalogs without a spectroscopic confirmation of their redshifts) of the Slug, for which we have optical LBT observation in 7 filters (Uspec, BBessel, VBessel, iSLOAN, gSLOAN, rSLOAN, zSLOAN). In addition, for the two QSOs, we used the photometric points reported in \citet{chen2021}, i.e., FUV, NUV, {\it J}, {\it H}, {\it Ks}, W1, W2, W3, W4, 450 $\mu$m, 850 $\mu$m, 2 mm, 3 mm and 1.4 GHz. We derived such quantities through multiwavelength spectral energy distribution (SED) fitting, using the \textsc{python} “Code Investigating GALaxy Emission" \citep[\texttt{CIGALE};][]{boquien2019cigale}. Firstly, we performed a match between the 36,880 detected source in the LBT observations and the prior catalogs of 69 LAEs and SMGs available for the Slug protocluster. For the resulting 26 sources with a counterpart, we performed two separate \texttt{CIGALE} runs. One for the 2/26 members with an X-ray counterparts (i.e., the central QSO UM287 and the companion QSO, compQSO), including the X-ray and radio modules, since data in both band are present. The other run, for the remaining 24/26 sources, was performed without including the aformentioned modules, even though we used a similar input parameters set for the other SED components. In particular, we adopted a typical delayed SFH, combined with the \citet{bruzual2003ssp} module for the stellar component; we selected the \citet{charlot2000dustatt} module for the the dust attenuation, along with the \citet{draine2014dustem} template for dust emission; finally, the AGN templates used are those from \cite{stalevski2012,stalevski2016} with the \texttt{skirtor2016} module. In Figure \ref{fig:SED_fitting} the results from the SED fitting of the two QSOs are shown. Both sources SED is dominated by the AGN across the electromagnetic spectrum, which in turn did not allow us to put significant constraints on the main physical quantities related to the host galaxy. Indeed, both the stellar masses and SFRs of the two QSOs are poorly constrained by the SED fitting. The AGN parameters are instead well constrained by the fit. The companion QSO (compQSO), has a large AGN fraction ($f_{\rm AGN} \sim 0.85$, defined as the ratio between the AGN and AGN + dust emission), and UM287 is dominated by the AGN contribution ($f_{\rm AGN} \sim 1$).
\par For the Slug members, we mainly investigated their stellar masses and SFRs (averaged in 100 Myrs timescale, as provided by the CIGALE output), both obtained from the \texttt{CIGALE} output. Figure \ref{fig:MS_members} shows the main sequence (MS) of galaxies for these sources, along with a MS at the same redshift from \citet{speagle2014MS}. We also color-coded the points following their AGN fractions. Most of the galaxies are below the MS, even though some systematics may be present due to the SED fitting templates. We compare our results with those from the XMM-SERVS \citep[][]{zou2022} at $2 < z < 2.5$, which are slightly below the MS, but overall in agreement with our data points at the same stellar masses.
CompQSO and UM287 are also displayed in the plot, with the former laying in the MS (log$(SFR/SFR_{\rm MS}) \sim -0.1$) and the latter being quite offsetted (log$(SFR/SFR_{\rm MS}) \sim -0.3$) . However, the results may be significantly affected by the large uncertainties on those values.

\end{appendix}

\end{document}